\documentclass[fleqn,usenatbib,12pt]{mnras}
 \usepackage{hyperref}
\usepackage{graphicx}	% Including figure files
\usepackage{subfigure}	% Including subfigure files
\usepackage{amsmath}	% advanced maths commands
\usepackage{amssymb}	% Extra maths symbols
\usepackage{bm}
\usepackage{ulem}
\usepackage{color}
\usepackage[T1]{fontenc}
\usepackage{ae,aecompl}
\usepackage{enumerate}
\newcommand{\oiii}{[O\,{\small III}]$\lambda$5007}
\newcommand{\nii}{[N\,{\small II}]$\lambda$6584}
\newcommand{\zoh}{12+\log \mathrm{(O/H)}}
\newcommand{\ab}{A_{\rm B}}
\newcommand{\av}{A_{\rm V}}
\newcommand{\afuv}{A_{\rm FUV}}

\defcitealias{Salim2020}{S20}	

\title[Systematic biases in deriving the $\av$--$\delta$ relation]{Systematic biases in determining dust attenuation curves through galaxy SED fitting}
\author[J. Qin et al.]{
Jianbo Qin,$^{1}$\thanks{jbqin@pmo.ac.cn}
Xian~Zhong~Zheng,$^{1,2}$\thanks{xzzheng@pmo.ac.cn}
Min~Fang,$^{1,2}$
Zhizheng~Pan,$^{1,2}$
Stijn Wuyts,$^{3}$
Yong~Shi,$^{4,5}$
\and 
Yingjie~Peng,$^{6}$
Valentino~Gonzalez,$^{7,8}$
Fuyan~Bian,$^{9}$
Jia-Sheng~Huang,$^{10,11}$
Qiu-Sheng~Gu,$^{4,5}$
\and 
Wenhao~Liu,$^{1,2}$
Qinghua~Tan,$^{1,2}$
Dong~Dong~Shi,$^{1}$
Jian~Ren,$^{1,2}$
Yuheng~Zhang,$^{1,2}$
Man~Qiao,$^{1,2}$
\and
Run~Wen,$^{1,2}$ 
and Shuang~Liu$^{1,2}$
\\
$^{1}$Purple Mountain Observatory, Chinese Academy of Sciences, 10 Yuanhua Road, Nanjing 210023, China\\
$^{2}$School of Astronomy and Space Sciences, University of Science and Technology of China, Hefei 230026, China\\
$^{3}$ Department of Physics, University of Bath, Claverton Down, Bath BA2 7AY, UK \\
$^{4}$School of Astronomy and Space Science, Nanjing University, Nanjing 210093,  China \\
$^{5}$Key Laboratory of Modern Astronomy and Astrophysics (Nanjing University), Ministry of Education, Nanjing 210093, China\\
$^{6}$Kavli Institute for Astronomy and Astrophysics, Peking University, 5 Yiheyuan Road, Beijing 100871, China\\
$^{7}$Chinese Academy of Sciences South America Centre for Astronomy, China-Chile Joint Centre for Astronomy, \\  \ \,Camino del Observatorio 1515, Las Condes, Chile\\
$^{8}$Centro de Astrofísica y Tecnologías Afines (CATA), Camino del Observatorio 1515, Las Condes, Santiago, Chile \\
$^{9}$European South Observatory, Alonso de Cordova 3107, Casilla 19001, Vitacura, Santiago 19, Chile\\
$^{10}$Chinese Academy of Sciences South America Center for Astronomy, National Astronomical Observatories, \\ \ \,Chinese Academy of Sciences, Beijing 100101, China \\
$^{11}$CAS Key Laboratory of Optical Astronomy, National Astronomical Observatories, Chinese Academy of Sciences, Beijing 100101, China}

\date{Accepted 2022 January 13. Received 2022 January 4; in original form 2021 November 23}
\pubyear{2022}
\begin{document}
\label{firstpage}
\pagerange{\pageref{firstpage}--\pageref{lastpage}}
\maketitle
\begin{abstract}
While the slope of the dust attenuation curve ($\delta$) is found to correlate with effective dust attenuation ($\av$) as obtained through spectral energy distribution (SED) fitting, it remains unknown how the fitting degeneracies shape this relation. We examine the degeneracy effects by fitting SEDs of a sample of local star-forming galaxies (SFGs) selected from the Galaxy And Mass Assembly survey, in conjunction with mock galaxy SEDs of known attenuation parameters. A well-designed declining starburst star formation history is adopted to generate model SED templates with intrinsic UV slope ($\beta_0$) spanning over a reasonably wide range. The best-fitting $\beta_0$ for our sample SFGs shows a wide coverage, dramatically differing from the limited range of $\beta_0<-2.2$ for a starburst of constant star formation. Our results show that strong degeneracies between $\beta_0$, $\delta$, and $\av$ in the SED fitting induce systematic biases leading to a false $\av$--$\delta$ correlation. Our simulation tests reveal that this relationship can be well reproduced even when a flat $\av$--$\delta$ relation is taken to build the input model galaxy SEDs. The variations in best-fitting $\delta$ are dominated by the fitting errors. We show that assuming a starburst with constant star formation in SED fitting will result in a steeper attenuation curve, smaller degeneracy errors, and a stronger $\av$--$\delta$ relation. Our findings confirm that the $\av$--$\delta$ relation obtained through SED fitting is likely driven by the systematic biases induced by the fitting degeneracies between $\beta_0$, $\delta$, and $\av$.           
\end{abstract}

\section{Introduction} \label{sec1}

Dust, which accounts for only a small fraction ($\sim$1\,per\,cent) of the interstellar medium (ISM) in star-forming galaxies (SFGs), can significantly change observables of the SFGs through absorbing and scattering stellar radiation \citep[][and references therein]{Galliano2018b}. The size distribution of dust grains and their chemical compounds regulate the degree of absorption and scattering across wavelength, described as the dust extinction curve \citep{Draine1984, Draine2003}. Interstellar dust containing more small grains yields a steeper extinction curve \citep{Weingartner2001, Draine2003, Hirashita2012, Asano2014, Hou2017, Aoyama2017}. In practice, extinction can be quantified via measuring light from a point source behind a dust screen. For an extended source like a galaxy, where dust and stars are mixed, dust attenuation is used to describe the deviation between the observed and the intrinsic stellar emission. The dust attenuation curve relies on not only the properties of the interstellar dust but also the geometry by which stars and dust are spatially distributed in the galaxy \citep{Witt1996, Witt2000, Seon2016, Narayanan2018, Lin2021}. Understanding the mechanisms/processes shaping the dust attenuation curve will provide key insights into the dust and structural evolution of galaxies.

It is well known that the shape of the dust extinction/attenuation curve varies significantly from one galaxy to another, including the Milky Way \citep[MW;][]{Fitzpatrick1986, Cardelli1989, Ferreras2021}, the Large Magellanic Cloud (LMC) and the Small Magellanic Cloud (SMC) \citep[][]{Pei1992, Gordon2003}, M31 \citep{Clayton2015}, as well as some galaxies in the nearby Universe \citep[e.g.][]{Calzetti2000, Gordon2003, Wild2011, Calzetti2021, Rezaee2021} and the distant Universe \citep[e.g.][]{Kriek2013,Reddy2015, Reddy2020, Shivaei2020, Kashino2021}. The variation is mostly attributed to the slope (or steepness) and the amplitude of the 2175\,\AA\ bump. The former is primarily governed by the size distribution of dust grains and the star-dust geometry \citep{Hirashita2012, Narayanan2018}, while the latter is probably caused by the graphite grains \citep[e.g.,][]{Mathis1994} or the polycyclic aromatic hydrocarbons \citep[PAH;][]{Weingartner2001, Draine2003}. The extinction/attenuation curves of the LMC and SMC are steeper than those of the MW and nearby starburst galaxies, and the 2175\,\AA\ bump is observed in the MW and LMC but not in the SMC and nearby starbursts.

In the past decade, many efforts have been devoted to addressing the relationships between the features of dust attenuation curves and galaxy properties. The slope of the dust attenuation curve was reported to depend on stellar mass \citep{Zeimann2015}, star formation rate \citep[SFR;][]{Teklu2020}, specific SFR \citep[$\rm sSFR=SFR/M_\ast$;][]{Reddy2015, Battisti2017b,Rezaee2021}, metallicity \citep{Battisti2017b,Shivaei2020}, as well as inclination \citep{Wild2011, Battisti2017a, Battisti2020}. However, some correlations remain controversial or inconsistent with each other, partially due to inconsistent datasets and methods used for drawing the conclusions.

Recently, an increasing number of studies determined the dust attenuation curves of galaxies by taking the slope of the curves as a free parameter in  modelling galaxy broad-band spectral energy distribution (SED). The SED fitting approach is applied to a large sample of galaxies. With this approach, an anti-correlation between the best-fitting slope of dust attenuation curves (e.g., UV-to-optical attenuation ratio $S\equiv\afuv/\av$) and dust column density (as approximately traced by $\av$) in SFGs has been established, in the sense that a flatter attenuation curve is linked with a higher $\av$ \citep[e.g.][]{Kriek2013, Arnouts2013, Salmon2016, Leja2017, Salim2018, Decleir2019, Battisti2020}. Theoretical investigations with radiative transfer models also predicted this $\av$--slope relation \citep{Witt2000, Chevallard2013, Narayanan2018, Trayford2020}. \citet[hereafter \citetalias{Salim2020}]{Salim2020} pointed out that $\av$ is the dominant driver of the attenuation curve slope, and at a fixed $\av$ (a proxy of dust column density) the slope does not show dependence on other galaxy parameters. If confirmed, this relation could play a key role in describing the variation of dust attenuation curves.

However, a fitting degeneracy exists between the two quantities involved in the $\av$--slope relation. Given that $\av$ is also used to define the attenuation curve slope (e.g., $A_{\rm FUV}/A_{\rm V}$), any errors in SED fitting that enlarge $\av$ would lead to a decrease of $\afuv/\av$, i.e. a flatter dust attenuation curve, when other parameters remain unchanged \citep{Salmon2016}. This issue was addressed in \citetalias{Salim2020}, showing that the degeneracy error from the given error ellipse (their Figure~9) is insignificant compared with the global relationship. By modelling the UV to NIR SEDs of star-forming regions in the SMC, \citet{Hagen2017} found a strong degeneracy between best-fitting $\av$ and attenuation curve slope (parameterized by $R_{\rm V}$\footnote{$R_{\rm V}=\av/(\ab-\av)$}). It is worth noting that there is a high degree of degeneracy between the star formation histories (SFHs) and dust attenuation curves adopted in SED fitting \citep{Calzetti2021}. The determination of the attenuation curve is sensitive to the adopted SFHs in the fitting \citep{Burgarella2005,Koprowski2020,Calzetti2021}. Considering the fitting degeneracy is largely unexplored, a thorough investigation is demanded to examine if the $\av$--slope relation is largely shaped by the fitting degeneracy. 

In this work, we aim to qualify the influence of fitting degeneracy on the well-established relation between $\av$ and attenuation curve slope. We use the observed data with secure measurements from the far-ultraviolet (FUV) to the far-infrared (FIR) to perform energy-balance fitting when modelling observed galaxy SEDs. In Section~\ref{sec2}, we briefly describe the galaxy sample and data used for our analysis. We introduce the parameters of our SED fitting in Section~\ref{sec3}. Section~\ref{sec4} presents our results of the fiducial fit configuration, while the results of a constant starburst fit configuration that is consistent with previous studies are given in Section~\ref{sec5} for comparison. We discuss our results in Section~\ref{sec6} and give a summary in Section~\ref{sec7}. A standard $\Lambda$CDM cosmology with $H_0=70$\,km$^{-1}$\,Mpc$^{-1}$, $\Omega _{\rm \Lambda}=0.7$, and $\Omega _{\rm m}=0.3$ and a \citet{Chabrier2003} Initial Mass Function (IMF) are adopted throughout the paper.

\section{SAMPLE AND DATA}\label{sec2}
\subsection{Sample selection}\label{sec2.1}

We carry out our investigation using the data from the Galaxy And Mass Assembly (GAMA) Survey, \footnote{\url{http://www.gama-survey.org/dr3} and all the GAMA value-added catalogues used in this work can be found in \url{http://www.gama-survey.org/dr3/data/cat/}}\citep{Driver2009,Driver2011}. GAMA is an optical comprehensive spectroscopic redshift survey over a 286\,deg$^2$ sky area divided into five different regions (with a limiting magnitude $r_{\rm petro}<19.8$\,mag) using the Anglo Australian Telescope’s AAOmega wide-field facility \citep{Driver2011,Hopkins2013,Liske2015}. 

The GAMA survey fields have extensive imaging data from the FUV to the FIR. These data come from different surveys: GALEX Medium Imaging Survey \citep[GALEX MIS;][]{Martin2005}; the Sloan Digital Sky Survey \citep[SDSS DR7;][]{Abazajian2009}, the VIsta Kilo-degree INfrared Galaxy survey \citep[VIKING;][]{deJong2013}; the Wide-field Infrared Survey Explorer \citep[WISE;][]{Wright2010} and the Herschel Astrophysical Terahertz Large Area Survey \citep[\textit{Herschel}-ATLAS;][]{Eales2010}. These data were collected and released to the public by \citet{Driver2016} through the GAMA Panchromatic Data Release. The software LAMBDAR\footnote{\url{http://gama-psi.icrar.org/LAMBDAR.php}} was used to measure fluxes from the  image data of 21 bands (FUV, NUV, $u$, $g$, $r$, $i$, $z$, $Z$, $Y$, $J$, $H$, $K_{\rm s}$, 3.4, 4.5, 12, 22, 100, 160, 250, 350, and 500\,$\micron$) \citep[see][for more details]{Wright2016}.  Elliptical apertures given by SExtractor were adopted to conduct photometry on the PSF-matched images, and corrections for the contamination by blended objects were employed.  In short, the GAMA datasets provide multi-band photometric catalogues for a large sample of spectroscopically-identified nearby galaxies with $r_{\rm petro}<19.8$\,mag. We use the datasets to measure the galaxy attenuation parameters through SED modelling based on the energy balance principle.

The original LAMBDAR photometric catalogues contain 219\,458 galaxies. Of them, 116\,261 sources have $r_{\rm petro}<19.8$\,mag and a reliable redshift measurement (${\rm nQ}\ge 3$). Both $r_{\rm petro}$ and nQ are taken from the TilingCat dataset. We limit targets over $0.07<z<0.2$. The lower limit of $z>0.07$ is chosen following \citet{Kewley2005} to ensure the GAMA/AAT 2$\arcsec$ fibre takes $>20$\,per\,cent of the total star light of a typical galaxy and minimize the potential differences between nuclear and global galaxy properties. The upper limit of $z<0.2$ is set to minimize evolutionary effects. There are 53\,182 GAMA galaxies in this redshift range. We also exclude faint sources with stellar masses less than $10^9$\,M$_\odot$. The stellar masses are from StellarMasses dataset, measured by fitting the observed $u$ to $K_{\rm s}$-band photometric data \citep{Taylor2011}. There are 52\,517 galaxies that meet our selection criteria.

Secure detections in multiple bands (including the FUV and the FIR) are needed for a robust determination of galaxy attenuation parameters through energy-balance SED fitting. Firstly, we select galaxies with a signal-to-noise ratio (S/N) greater than three in all bands from FUV to $K_{\rm s}$,\footnote{ Here the VISTA/VIRCAM $Z$ band is not included.} leaving 17\,796 out of 52\,517 galaxies. Galactic extinction was corrected for all fluxes from FUV to $K_{\rm s}$ using the \citet{Schlegel1998} MW dust maps \citep{Wright2016}. To securely measure dust emission, we focus on the target galaxies with good detections (${\rm S/N}>3$) in at least one of five \textit{Herschel} PACS and SPIRE bands, leaving 8\,531 galaxies with an FIR detection rate of 48\,per\,cent. It is worth mentioning that the S/N cut in the FIR likely biases  our sample selection towards galaxies being more dusty. Since we focus on investigating the parametrized relations and the sample still retain a dynamical range of nearly one order of magnitude in $\av$ ([0.1, 1.5]\,mag), the FIR selection cut will not significantly affect our results.  

We also make use of WISE 12\,$\micron$ and 22\,$\micron$ data, if available, to improve the measure of total infrared (IR) luminosity (8--1000\,$\micron$). Most of our galaxies detected by \textit{Herschel} have secure detections in either WISE 12\,$\micron$ or 22\,$\micron$ (${\rm S/N}>3$, the detection rate is $\sim$83\,per\,cent). We measure the IR luminosity via best-fitting the observed IR data points with the IR SED templates from the dust radiation model by \citet{Draine2007}. We let the PAH fraction vary from 0.47 to 4.58, the minimum radiation field $U_{\rm min}=[0.1, 25]$, the maximum radiation field $U_{\rm max}=[10^3, 10^6]$, and the fraction illuminated from $U_{\rm min}$ to $U_{\rm max}$ is [0, 1]. If the IR bands which fall below 3\,$\sigma$ do have valid flux and error measurements, they are also included to constrain the IR luminosity measurements. For the bands without detections, the upper limits are used in the fitting. The typical error is $\sim0.1$\,dex for the measured IR luminosities. 

On the other hand, AGN activity may play a role in heating up the dust and thus contributing to the IR emission \citep{Mullaney2011, Kirkpatrick2015}, which increases the uncertainties in the modelling of SEDs. We select `star-forming' galaxies without signs of nuclear activity based on the BPT diagram \citep{Baldwin1981}. This requires good measurements of the emission line fluxes including H$\alpha$, [N\,{\sc ii}], H$\beta$, and [O\,{\sc iii}]. These line fluxes are taken from the SpecLine dataset \citep{Gordon2017}. All lines have been measured by fitting the spectral line with a single Gaussian function. Following \citet{Salim2018}, we require ${\rm S/N}>10$ for H$\alpha$ and ${\rm S/N}>2$ for the remaining three lines. As pointed out, if we care about the line ratios (such as \nii/H$\alpha$ and \oiii/H$\beta$ used in the BPT diagram as well as the metallicity determination), the usual cut ${\rm S/N}>3$ is too strict. When the S/N cut is relaxed, some sources with weak emission lines can be picked up and the sample size may increase. Finally, there are 2\,764 sources satisfying our S/N cuts and classified as SFGs following the criteria given by \citet{Kauffmann2003}.   

The gas-phase metallicity, parameterized by Oxygen abundance O/H, is estimated from the N2 method using the formula given by \citet{Pettini2004} as
\begin{equation}\label{eq_pp04}
\zoh = 9.37 + 2.03\times {\rm N2} + 1.26\times {\rm N2}^2 + 0.32\times {\rm N2}^{3},
\end{equation}
where $\rm N2 = \log$(\nii/H$\alpha$). The N2 method is adopted since the emission lines in N2 are also used to identify SFGs with the BPT diagram. Our metallicity measurements can be carried out for the entire sample without introducing any additional selection criteria. Equation~\ref{eq_pp04} is valid over $-2.5<{\rm N2}<-0.3$, corresponding to $7.17<\zoh<8.86$ \citep{Pettini2004}. In our analysis, we exclude 9 galaxies with N2 out of this range. This will not affect our results.

The half-light radius ($R_{\rm e}$) and axial ratio (b/a) are taken from \citet{Kelvin2012}, who presented a single-S\'{e}rsic two-dimensional model fits to SDSS images for 167\,600 galaxies in the GAMA data base. In this work, we adopt the $r$-band half-light radius and axial ratio. Using structural parameters of other SDSS bands does not alter our conclusions. We consider the best-fitting reduced chi-square ($\chi_{\rm r}^2\equiv\chi^2/N_{\rm dof}$, where $N_{\rm dof}$ is the number of degrees of freedom) in the range $0.5<\chi_{\rm r}^2<1.5$ as reliable measurements for morphological and structural parameters. We select disc galaxies with S\'{e}rsic index less than two. For galaxies with a S\'{e}rsic index greater than two, their shapes tend to be more spheroidal \citep{Padilla2008}, and the axial ratio is no longer a good probe of the inclination. We also exclude 19 extreme edge-on galaxies with axial ratio less than 0.15 from our sample, as the scale heights of these galaxies will bias the linkage between inclination and axial ratio \citep{Guthrie1992}. Our final sample contains 2\,291 SFGs over $0.07<z<0.2$ with secure detections in multiple bands, as well as the gas-phase metallicity and structural parameters.

\begin{figure*}%[htb]
\centering
\includegraphics[width=0.9\textwidth]{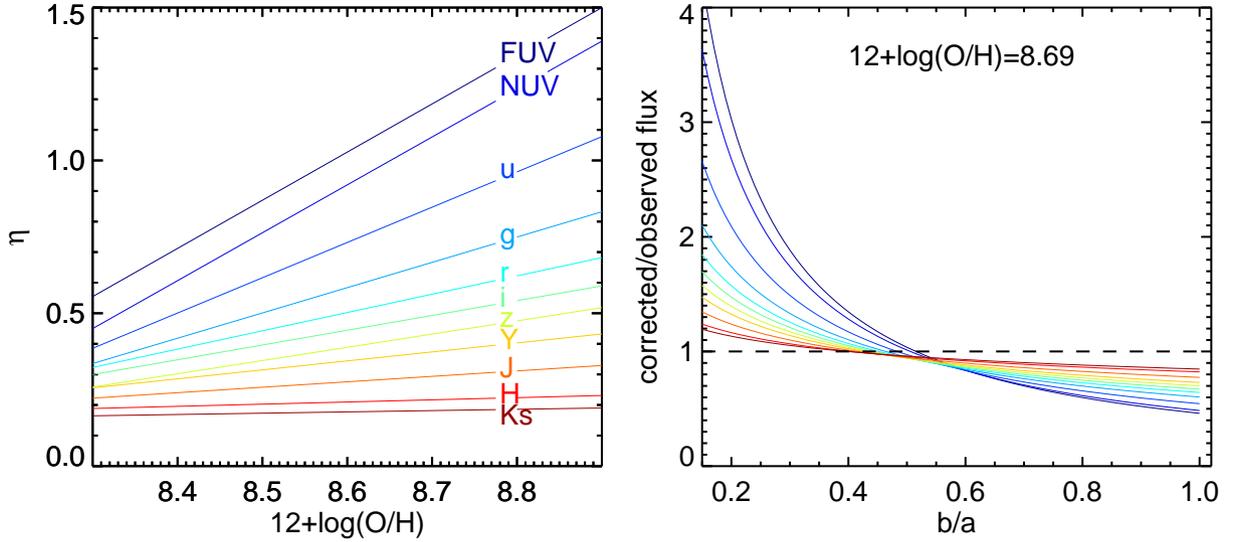}
\caption{Left: Best-fitting power-law index $\eta$ quantifying the inclination-dependent luminosity (Eq.\ \ref{eq_llam_llamf}) as a function of metallicity. The solid lines from top to bottom represent the relations from FUV to $K_{\rm s}$. Right: The corrected-to-observed flux ratio as a function of axial ratio for galaxies with Solar metallicity. These lines are colour-coded by bands in the same way as in the left panel. }
\label{fig_fluxcor}
\end{figure*}

\subsection{Flux correction for inclination-dependent  inhomogeneity}\label{sec2.2}

When calculating the total luminosity of a disc galaxy over a solid angle of 4$\pi$ radians, we often assume that its radiation is homogeneous in all directions. However, the UV and optical radiation are attenuated by dust that is mostly distributed in the disc and thus the UV/optical radiation is no longer uniform in all directions.  The observed UV/optical flux is dependent on inclination.  In contrast, the IR radiation is nearly free from dust attenuation and thus evenly emits in all directions.  As a consequence, the UV/optical flux is underestimated (overestimated) for the edge-on (face-on) galaxies. This induces a tension in balancing  energy between the UV/optical and the IR in SED fitting. This tension might result in artificial effects on dust attenuation \citep[see][and references therein]{Doore2021}. More importantly, the bias in the observed fluxes is strongly wavelength-dependent, and consequently influences the dust attenuation curve.  In our analysis, the observed fluxes (and luminosities) in FUV to $K_{\rm s}$ are corrected for the inclination-induced bias and the `corrected' values (i.e. the fluxes averaged over the 4$\pi$ solid angle) are used to construct the observed SEDs.

Consider a galaxy with a brightness distribution of $L_{\lambda}(\phi,\theta)$, where $\phi$ and $\theta$ are the azimuth angle ([0, $2\pi$]) and polar angle([$-\pi/2$, $\pi/2$]), respectively. Then the corrected luminosity can be obtained as
\begin{align} \label{eq_llam_thetai}
L_{\lambda,{\rm corrected}}&=\frac{1}{4\pi}\int_0^{4\pi}L_{\lambda}(\phi,\theta) d\Omega,  \nonumber \\
                   &=\frac{1}{4\pi}\int_0^{2\pi}d\phi \int_{-\frac{\pi}{2}}^{\frac{\pi}{2}} L_{\lambda }(\phi,\theta) \sin(\theta) d\theta. 
\end{align}
For a disc galaxy that is rotationally symmetrical along the $\phi$ direction and symmetrical along the $\theta$ direction, then Equation~\ref{eq_llam_thetai} can be written as
\begin{align} \label{eq_llam_ar}
 L_{\lambda,{\rm corrected}}&=\int_{0}^{\frac{\pi}{2}} L_{\lambda}(\theta) \sin(\theta) d\theta, \nonumber \\
                    &=\int_{0}^{1} L_{\lambda}[cos(\theta)] d[cos(\theta)],  \nonumber \\
                    &\approx\int_{0}^{1} L_{\lambda}({\rm b/a})d({\rm b/a}). 
\end{align}
The random projection of a disc galaxy in $4\pi$ results in a distribution of galaxy inclination over [0, $\pi$/2] or $\rm b/a=[0, 1]$. The corrected luminosity can be calculated by integrating the luminosity distribution over the range of b/a.  

\citet{Qin2019a} found there is a tight power-law relation between ${\rm IRX}=L_{\rm IR}/L_{\rm UV}$ and b/a. Since $L_{\rm IR}$ is not affected by inclination, it is equivalent to a power-law relation between b/a and $L_{\rm UV}$. Hence, we assume that the axial ratio and luminosity of each band satisfy a power-law relation $L_{\lambda}\propto({\rm b/a})^{\eta}$. If the luminosity of a galaxy viewed face-on (${\rm b/a=1}$) is $L_{\lambda}^{f}$, then the luminosity at any b/a should be
\begin{equation}\label{eq_llam_llamf}
L_{ \lambda}=L_{ \lambda}^{f}\times({\rm b/a})^\eta.
\end{equation}
Considering the redshift range of our sample is rather narrow ($0.07<z<0.2$), we ignore the band-shifting effect. We then substitute Equation~\ref{eq_llam_llamf} into the Equation~\ref{eq_llam_ar}, and have
\begin{align} \label{eq_llam_final}
L_{\lambda,{\rm corrected}}&=\int _0^1 L_{\lambda}^{f}({\rm b/a})^\eta d({\rm b/a}) \nonumber \\ 
                   &=L_{\lambda}^{f}/(1+\eta),  \nonumber \\ 
                   &=\frac{L_{\lambda}}{(1+\eta)({\rm b/a})^\eta}. 
\end{align}
With the observed luminosity, axial ratio and the power-law index $\eta$, the corrected luminosity for each band can be derived accordingly. \citet{Qin2019a} developed a novel method to obtain the intrinsic relation between axial ratio and IRX.  The dependence of IRX on IR luminosity, metallicity, galaxy size and axial ratio were quantified by minimizing the dispersion of IRX in fitting the data points with multiple power-law functions.  Given that IR luminosity is not dependent on galaxy inclination, the correlation between b/a and IRX is governed by the correlation between b/a and $L_{\rm UV}$. 

Similarly, we use the same galaxy parameters to minimize the scatter of the luminosity in each band as,
 \begin{equation}\label{eq_llam_paras}
L_{\lambda}=10^\alpha\,(\frac{L_{\rm IR}}{10^{10}\,{\rm L}_\odot})^{\beta}\,(\frac{R_{\rm e }}{\rm kpc})^{-\gamma}\,(\rm b/a)^{-\eta},
\end{equation}
where $\alpha$, $\beta$, $\gamma$, and $\eta$ are power-law exponents respectively. \citet{Qin2019a} found that these indices depend on the gas-phase metallicity, as
\begin{equation}\label{eq_paras_zgas}
 X=c_X \log(Z/\rm Z_\odot)+d_X,
\end{equation}
$X$ represents $\alpha$, $\beta$, $\gamma$, or $\eta$, and $c_X$ and $d_X$ are their respective coefficients. By best fitting the luminosity in each band, we obtain $\eta$.

Figure~\ref{fig_fluxcor} shows the best-fitting power-law index $\eta$ as a function of metallicity. $\eta$ decreases from FUV to $K_{\rm s}$, indicating that radiation at a shorter wavelength is more affected by the increasing dust attenuation (and decreasing b/a). We find the metal-poor galaxies have smaller $\eta$, i.e. flatter inclination-luminosity relation. This is consistent with the results presented in \citet{Qin2019a}. They pointed out that the low-metallicity SFGs are usually less massive and tend to be more spheroidal in morphology, and the axial ratio is no longer decided by inclination.  At increasing wavelength, e.g. from FUV to $K_{\rm s}$, the dependence of luminosity on metallicity becomes gradually weaker.  

We use Equation~\ref{eq_llam_final} to derive the corrected flux from the measured flux in a given band. The flux error is also adjusted to match the conversion. The right panel of Figure~\ref{fig_fluxcor} shows the corrected-to-observed flux ratio as a function of axial ratio for galaxies with Solar metallicity of $12+\log (\rm O/H)=8.69$. We can see that for edge-on galaxies (i.e., b/a = 0.2),  the corrected-to-observed flux ratio is  3 and 1.2 in FUV and $K_{\rm s}$, respectively. And for face-on galaxies (i.e., b/a = 1), the ratio becomes 0.5 and 0.8, respectively. The corrected flux approximately equals the observed flux when the axial ratio is between 0.4 and 0.6, varying with metallicity and wavelength. In other words, the correction is equivalent to rotating all galaxies to b/a$\sim 0.4-0.6$. It is clear that the inclination effect is wavelength-dependent, and it may affect the further determination of the attenuation curve. We use the corrections given by Equation~\ref{eq_llam_final} to convert the observed fluxes into the corrected fluxes and perform further analyses with data corrected for the inclination effect. 

By doing this correction, we obtain SEDs satisfying the energy balance. One caveat is that the correction still suffers from some uncertainties. For example, we use the optical-band structure parameters (i.e., $r$-band) for the UV fluxes  without corrections for the colour-gradient effects. We suspect that these uncertainties are marginal and our results are not significantly affected. In fact, our conclusions are insensitive to the inclination corrections. We verify that our main conclusions are not influenced even if using the inclination-uncorrected data.

\begin{figure}
\centering 
\includegraphics[width=0.9\columnwidth]{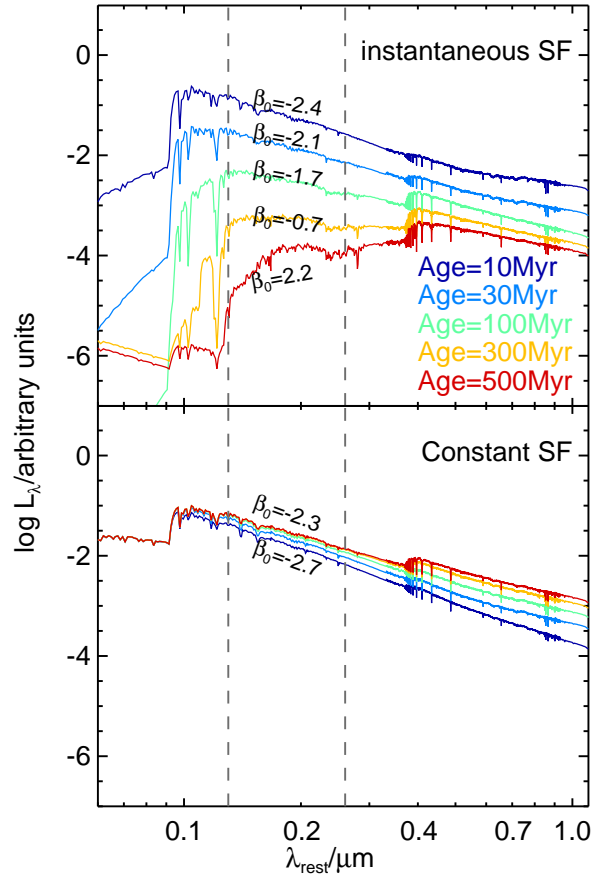}
\caption{Top: The intrinsic SEDs of galaxy stellar populations from an instantaneous star formation at the ages of $10-500$\,Myr for solar metallicity. The UV slope $\beta_0$, defined as $L_{\rm \lambda}\propto\lambda^{\beta_0}$ in the UV, is labelled for each SED. The two dashed lines mark the wavelength range for determining the UV slope. Bottom: Similar to the top panel but for a constant star formation.} 
\label{fig_sfh_delay}
\end{figure}

\section{Fitting galaxy SEDs with CIGALE}\label{sec3}

We analyse the observed SEDs of our sample galaxies using the Code Investigating GALaxy Emission\footnote{\url{https://CIGALE.lam.fr}}\citep[CIGALE;][]{Noll2009,Boquien2019}. The basic idea of CIGALE is that the total energy radiated by the dust in the IR equals the total energy absorbed by the dust in the UV/optical. CIGALE combines a library of single stellar populations (SSP) and variable attenuation curves with SFH models to generate a large number of grid SED models to fit the observed data. The modelled SEDs are integrated into a set of filters and compared directly to the observations. The observations are assigned with an extra 10\,per cent uncertainty (done by CIGALE itself) to account for the uncertainties from the models themselves \citep[see][]{Noll2009}. The output parameters are measured with the Bayesian likelihood statistics method based on the probability distribution functions (PDFs). The best-fitting parameters and the corresponding uncertainties are the likelihood-weighted mean\footnote{$\overline{X}=\sum\limits_{i}(X_i P_i)/\sum\limits_{i}P_i$} and the standard deviation of all models. Details of CIGALE can be found in \citet{Boquien2019}. 

We model the observed galaxy SEDs with the following components: stellar emission, nebular lines from ionized gas, dust attenuation and dust emission. Here the AGN component is not taken into account because our sample contains only SFGs. Following \citet{Salim2018} and \citet{Decleir2019}, we fit the UV-to-NIR part of an observed SED, and meanwhile, total IR luminosity derived from observed IR data points is taken as an additional IR `data point' to balance the dust absorption. The determination of the total IR luminosity is described in Section~\ref{sec2.1}.

\begin{figure*}
\centering 
\includegraphics[width=0.9\textwidth]{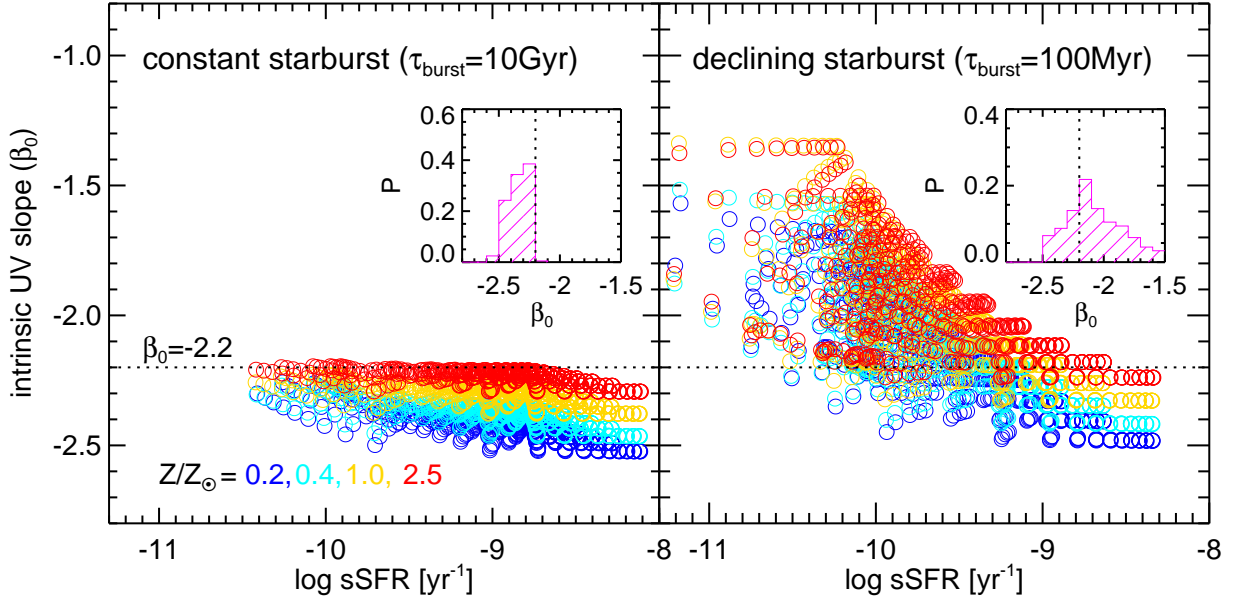}
\caption{The intrinsic UV slope $\beta_0$  as a function of sSFR for our model SEDs.  The models are generated with two starburst settings: one is a starburst of constant star formation (left) and the other is a declining starburst (right). The symbols are colour-coded with stellar metallicity 0.2 (blue), 0.4 (cyan), 1.0 (gold), and 2.5 (red) times Z$_\odot$. Note that $\beta_0$ decreases with sSFR and increases with metallicity. The dotted lines in both panels mark $\beta_0=-2.2$. The inner panels show the $\beta_0$ histogram of our model SEDs. The vertical dotted lines mark $\beta_0=-2.2$. } 
\label{fig_ssfr_beta0}
\end{figure*}

\begin{table*}
 \centering
 \caption{Modules and input parameters with CIGALE for generating our model galaxy SEDs. The two configurations have different $\tau_{\rm burst}$.  \label{tab1}}
\scalebox{1}{
 \begin{tabular}{lll}
  \hline\hline
  Module                               &Parameter                          &Value\\
  \hline
  \texttt{sfh2exp}                     &\texttt{age\_main} (Myr)           &11000\\
                                       &\texttt{tau\_main} (Myr)           &3000, 5000, 7000, 9000, 11000\\
                                       &\texttt{age\_burst} (Myr)          &100, 150, 200, 250, 300, 350, 400, 450, 500\\
                                       &\texttt{tau\_burst}                &I. 100\,Myr (fiducial declining starburst fit)\\
                                       &                                   &II. 10\,Gyr (constant starburst fit)\\
                                       &\texttt{f\_burst}                  &0.01, 0.05, 0.10, 0.15, 0.20, 0.25, 0.30, 0.35, \\
                                       &                                   &0.40, 0.45, 0.50 \\\hline
  \texttt{bc03}                        &\texttt{imf}                       &1 (Chabrier)\\
                                       &\texttt{metallicity}               &0.004, 0.008, 0.02, 0.05\\\hline
  \texttt{nebular}                     &\texttt{logU}                      &$-3.0$\\
                                       &\texttt{f\_esc}                    &0.0\\
                                       &\texttt{f\_dust}                   &0.0\\
                                       &\texttt{lines\_width} (km s$^{-1}$)&300\\\hline
  \texttt{dustatt\_modified}           &\texttt{E\_BV\_lines}(mag)         &0.10, 0.15, 0.20, 0.25, 0.30, 0.35, 0.40, 0.45,\\ 
   \texttt{\_starburst}                &                                   &0.50, 0.55, 0.60, 0.65, 0.70, 0.75, 0.80\\
                                       &\texttt{E\_BV\_factor}             &0.44\\
                                       &\texttt{uv\_bump\_amplitude}       &0 (no bump)\\
                                       &\texttt{powerlaw\_slope}           &$-$1.4, $-$1.2, $-$1.0, $-$0.8, $-$0.6, $-$0.4, $-$0.2, \\
                                       &                                   & 0.0, 0.2, 0.4, 0.6\\
                                       &\texttt{Ext\_law\_emission\_lines} &1 (Milky Way)\\
                                       &\texttt{Rv}                        &3.1\\ \hline
  \texttt{dale2014}                    &\texttt{alpha}                     &2.0\\
                                       &\texttt{f\_AGN}                    &0\\ \hline
                                       \hline
   \end{tabular}}
\end{table*}

\subsection{Star formation histories}\label{sec3.1}

To build a stellar composition, we use the BC03 stellar population synthesis models \citep{Bruzual2003} with a \citet{Chabrier2003}  initial mass function. The BC03 models are made with six stellar metallicities. We use four of them from 0.2 to 2.5\,Z$_\odot$, which is adequate for most galaxies in the local universe \citep{Gallazzi2005}. Following \citet{Salim2016,Salim2018}, we use the two-component exponential models (sfh2exp) offered by CIGALE to set SFHs. It consists of an exponential main (old) component and a recent exponential starburst. Both components are parameterized by the age (t) and e-folding time ($\tau$).  The age of the main component ($t_{\rm main}$) is fixed at 11\,Gyr, and the e-folding time ($\tau_{\rm main}$) varies from 3 to 11\,Gyr in a step of 2\,Gyr. The choice of a fixed old stellar age of the main component is to avoid the potential risk of returning unrealistic young stellar ages in the fitting\citep[e.g.,][]{Salim2016,Salim2018,Decleir2019,Boquien2019,Nersesian2019}.   
 
For the starburst component, we notice that a starburst with constant star formation was often adopted in previous studies \citep{Buat2011a, Salim2018, Salim2019, Decleir2019, Salim2020}. The constant starburst is considered as a convenient approximation in SED modelling, but it has a drawback in generating templates with representative intrinsic UV colours (no dust). \citet{Koprowski2020} showed that a constant starburst creates spectral templates of different ages with similar intrinsic UV colours, being less sensitive to the life time of the burst \citep[see also][]{Calzetti2001,Mao2012}. In Figure~\ref{fig_sfh_delay} we compare the spectral templates in two extreme cases: one an instantaneous starburst (or a single stellar population) and the other a constant starburst. The intrinsic UV slope $\beta_0$, defined as the index in the power-law relationship $L_\lambda \propto \lambda^{\beta_0}$ in UV, is measured for each of the intrinsic SED template (no dust) following the methodology of \citet{Calzetti1994}. Here metallicity is fixed to Solar. One can see that by increasing age  over $10-500$\,Myr,  $\beta_0$ changes dramatically from $-$2.4 to 2.2 for the instantaneous starburst, while $\beta_0$ mildly increases from $-2.7$ to $-2.3$ for the constant starburst. This is because the galaxy population is constantly replenished by the youngest stellar populations that dominate the UV radiation and result in an approximately constant UV colour \citep{Koprowski2020,Calzetti2021}. 

For our two-component SFH prescription, the UV colour is decided by not only the starburst component but also (partially) the main component. To model the intrinsic SED of a galaxy, we let the age of the starburst ($t_{\rm burst}$) vary from 100\,Myr to 500\,Myr in a step of 50\,Myr. We set up two types of starbursts, having an e-folding time ($\tau_{\rm burst}$) of 100\,Myr (declining) and 10\,Gyr (constant). The mass fraction of the starburst component varies from 1 to 50\,per\,cent. Combined together,  2376 SED models were generated with different SFHs at each fixed $\tau_{\rm burst}$. The modules and input parameters used to generate model SEDs with CIGALE are presented  in Table~\ref{tab1}. We also include nebular emission in estimating  UV slope, being consistent with the settings in our SED fitting.  More details of modelling nebular emission are described in Section~\ref{sec3.3}.

Figure~\ref{fig_ssfr_beta0} shows $\beta_0$  as a function of sSFR for model SEDs generated with the two types of starbursts. We can see that both starburst settings produce a reasonable range of sSFR for local SFGs \citep{Guo2015}. We find that $\beta_0$ decreases with sSFR, suggesting that younger stellar populations have bluer SEDs in the UV. At given sSFR $\beta_0$ increases with stellar metallicity, saying that metal-rich galaxies have higher $\beta_0$ (i.e. redder UV colour). The intrinsic linkage between $\beta_0$, sSFR, and metallicity is consistent with that in \citet{Salim2019}. For a constant starburst, the $\beta_0$ of the generated model SEDs spread over a small range at $<-2.5\beta_0<-2.2$ for all four metallicities; For a declining starburst, the model SEDs spread over a wide range of $-2.5<\beta_0<-1.3$. This difference of model spreads between the left and right panels is consistent with that given in Figure~\ref{fig_sfh_delay}. Again, the constant starburst keeps the UV colour of model SEDs barely changed. Meanwhile, the declining starburst allows a significant fraction of intermediate-age stellar populations to create model SEDs with redder UV colours. 

Accumulating evidence from both the observational and theoretical sides shows that galaxies are characterized by bursty and episodic SFHs governed by non-smoothing processes \citep[e.g.][]{Sparre2017,Iyer2020}. The bursty SFHs cause the intrinsic UV slope ($\beta_0$) of the stellar populations to vary over a wide range \citep[e.g.][]{Boquien2012, Battisti2016, Schulz2020,Calzetti2021}. This will be further discussed in Section~\ref{sec6.1}. Considering that the dust attenuation (curve) is most sensitive to the UV radiation \citep[][]{Draine2003, Galliano2018b, Narayanan2018, Butler2021}, a steep slope $\beta_0<-2.2$ for all model SED templates will bias the dust attenuation curve estimated from SED fitting. For example, fitting the observed data with model SEDs of a bluer intrinsic UV colour will result in a steeper attenuation curve according to the degeneracy between SFHs and attenuation curves \citep{Calzetti2021}. In our SED fitting, we take the model SEDs with SFHs of a declining starburst ($\tau_{\rm burst}=100$\,Myr) to conduct our fiducial `declining starburst fit', as it scans a reasonably wide range of $\beta_0$. For comparison, we also perform SED fitting using model SEDs with SFHs of a constant starburst (i.e. $\tau_{\rm burst}=10$\,Gyr) but keep other parameters unchanged. The results are referred to as `constant starburst fit'. The configurations for the two fittings are listed in Table~\ref{tab1}.

\subsection{Dust attenuation laws}\label{sec3.2}

We adopt the modified \citet{Calzetti2000} Law to describe the dust attenuation curve. Specifically, modifying the \citet{Calzetti2000} attenuation curve with a slope deviation and the 2175\,\AA\ absorption feature \citep{Noll2009} as 
\begin{equation} \label{eq_modified_calzetti_law}
A(\lambda)={\rm E(B-V)}\left[k(\lambda)\left(\frac{\lambda}{\lambda_{\rm V}}\right)^{\delta}+D(\lambda)\right],
\end{equation}
where $\lambda_{\rm V}=0.55\,\mu$m, E(B$-$V) is the colour excess defined as E(B$-$V)$\equiv A_{\rm B}-A_{\rm V}$, $k(\lambda)$ is the \citet{Calzetti2000} attenuation curve normalized on E(B$-$V), $\delta$ is the deviate power-law slope, and $D(\lambda)$ is the 2175\,\AA\ absorption bump (or UV bump).  If $\delta=0$ and no bump is included, Eq.\ \ref{eq_modified_calzetti_law} reverts backs to the original Calzetti attenuation curve; and for the Milky Way, it roughly corresponds to $\delta\approx 0.15$ with a bump strength $E_{\rm b}$ (normalization of $D(\lambda)$) of $\approx 3$. Constraining the strength of the 2175\,\AA\ bump requires multiple-band photometry or spectroscopy in the UV \citep[e.g.][]{Calzetti1994,Buat2011b,Kriek2013,Decleir2019,Kashino2021}. There are only two broad UV bands included in our dataset and the bump is thus poorly constrained \citep[see also][]{Salim2018}. Here we fix $E_{\rm b}=0$, i.e. no 2175\,\AA\ bump. We stress that setting the 2175\,\AA\ bump as a free parameter or fixing it do not alter our conclusions. CIGALE allows us to separate the young and old stellar populations. The young stellar populations are mostly in the star-forming regions and the old populations are mostly mixed with the diffuse ISM. The former suffers higher attenuation than the latter \citep[][]{Calzetti1994, Charlot2000, Wild2011, Qin2019b, Lin2020, Li2021}. \cite{Calzetti1994} found  that stellar continuum and nebular lines exhibit a different degree of dust attenuation, giving an E(B$-$V) ratio of $\sim$0.44 on average. We adopt this value in our SED fitting and use the age of 10\,Myr as the boundary to distinguish young and old stars. Here both old and young stellar populations share the same attenuation law but have different E(B$-$V).

\subsection{The nebular emission lines}\label{sec3.3}

CIGALE is able to deal with ionized gas radiation (i.e. emission lines) in the model, which has a moderate effect on broadband fluxes and colours of galaxies\citep[e.g.,][]{Salim2016,Yuan2019}. We found that the inclusion of emission lines will moderately increase the quality of SED fitting (decreases of reduced $\chi^2$). Here we use the \citet{Inoue2011} ionized gas radiation model to simulate the emission lines in galaxies. The model is based on CLOUDY 13.01\citep{Ferland1998,Ferland2013}. Following \citet{Boquien2019}, we set the ionization parameter to $\log U=-3.0$ and the fraction of Lyman continuum photons absorbed by dust to $f_{\rm dust}=0$. We find that using a larger or smaller value has no effect on the conclusions. These emission lines are attenuated with a fixed Milky Way extinction curve, while their E(B$-$V) is consistent with that of young stars. 

All the modules and parameters are summarized in Table \ref{tab1}. Combined all possible values for all parameters, a total of 392\,040 sub-models are generated for each redshift increased by 0.01 over the sample redshift range.   CIGALE is run under two configurations: the fiducial `declining starburst fit' and the `constant starburst fit'. The latter is set to be consistent with previous studies, including \citetalias{Salim2020}. The main difference between the two configurations is the $\beta_0$ coverage (see Figure~\ref{fig_ssfr_beta0}). We show the results of the fiducial declining starburst fit in Section~\ref{sec4} and then show the results of the constant starburst fit for comparison in Section~\ref{sec5}.

\begin{figure*}%[htb!]
\centering
\includegraphics[width=0.8\textwidth]{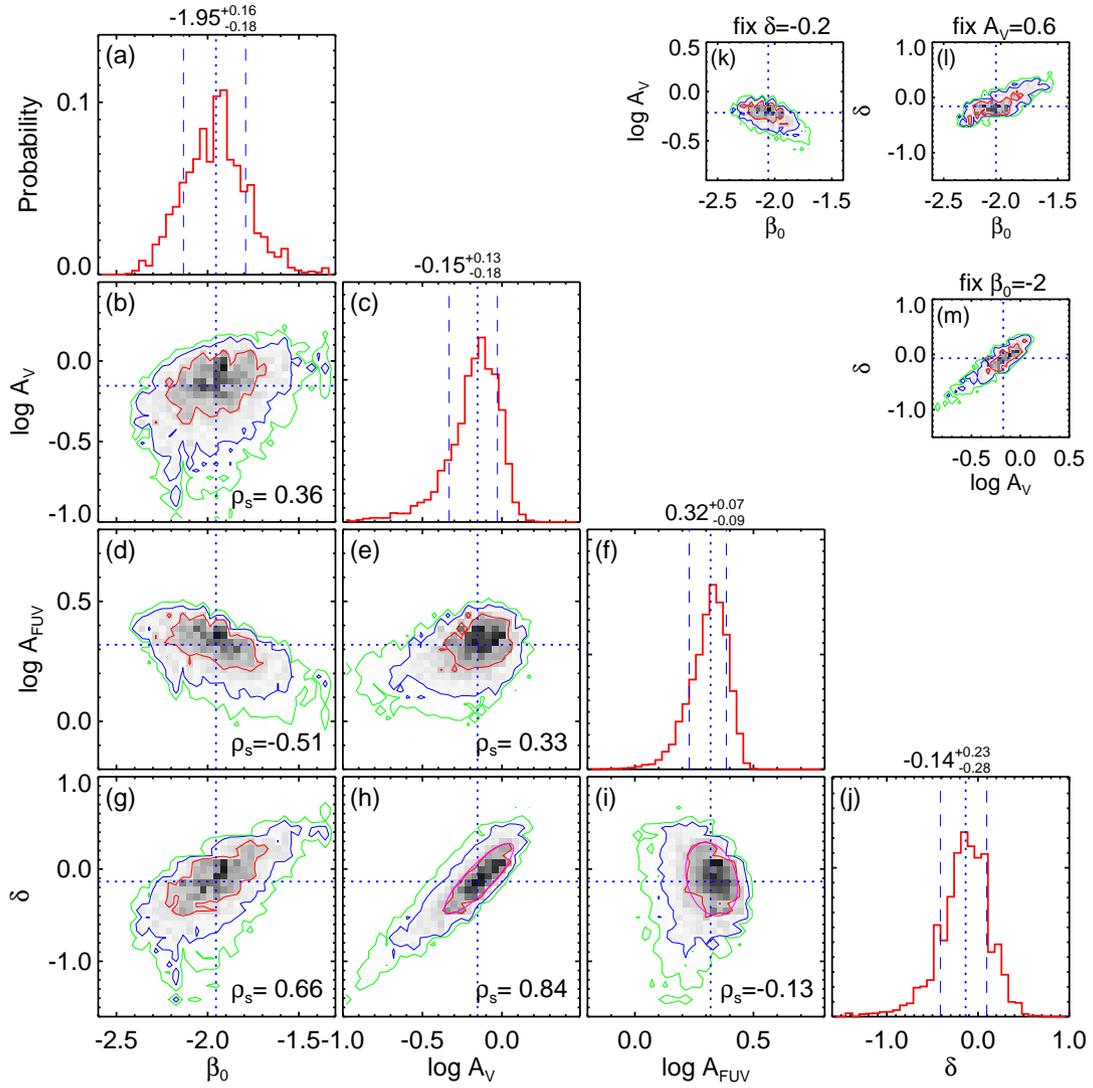}
\caption{The posterior probability distributions of $\beta_0$, $\av$, $\afuv$, and $\delta$ from our SED fitting of a typical galaxy. Here the declining starburst is adopted in building SFHs and model SEDs. In each panel, the blue dotted lines mark the 50th percentile values (slightly differ from the likelihood-weighted mean values). In each histogram plot, the 16th and 84th percentiles are marked with the dashed lines. The red, blue and green contours enclose 68, 95, and 99\,per\,cent of probability (i.e. 1, 2, and 3\,$\sigma$), respectively.  The magenta ellipses in panels (h) and (i) best fit the 1\,$\sigma$ contours. The Spearman's rank correlation coefficient $\rho_{\rm s}$ is also presented in each panel. The top-right three panels (k, l, and m) show the probability distributions if fixing one parameter. }
\label{fig_bayes_matrix}
\end{figure*}

\section{Results from the fiducial declining starburst fit} \label{sec4}

We show in this section the results of fitting the observed SEDs of local SFGs with CIGALE for our fiducial declining starburst fit. We firstly examine the possible degeneracies of output parameters, including $\afuv$, $\av$, $\delta$, and $\beta_0$ in Section~\ref{sec4.1}. We then show the dependence of best-fitting $\delta$ on $\av$ (and $\afuv$) of our sample galaxies in Section~\ref{sec4.2}. Finally, in Section~\ref{sec4.3} we evaluate the effect of fitting degeneracy on $\av$--$\delta$ relation using simulated data.  
 
\begin{figure*}%[htb!]
\centering
\includegraphics[width=0.9\columnwidth]{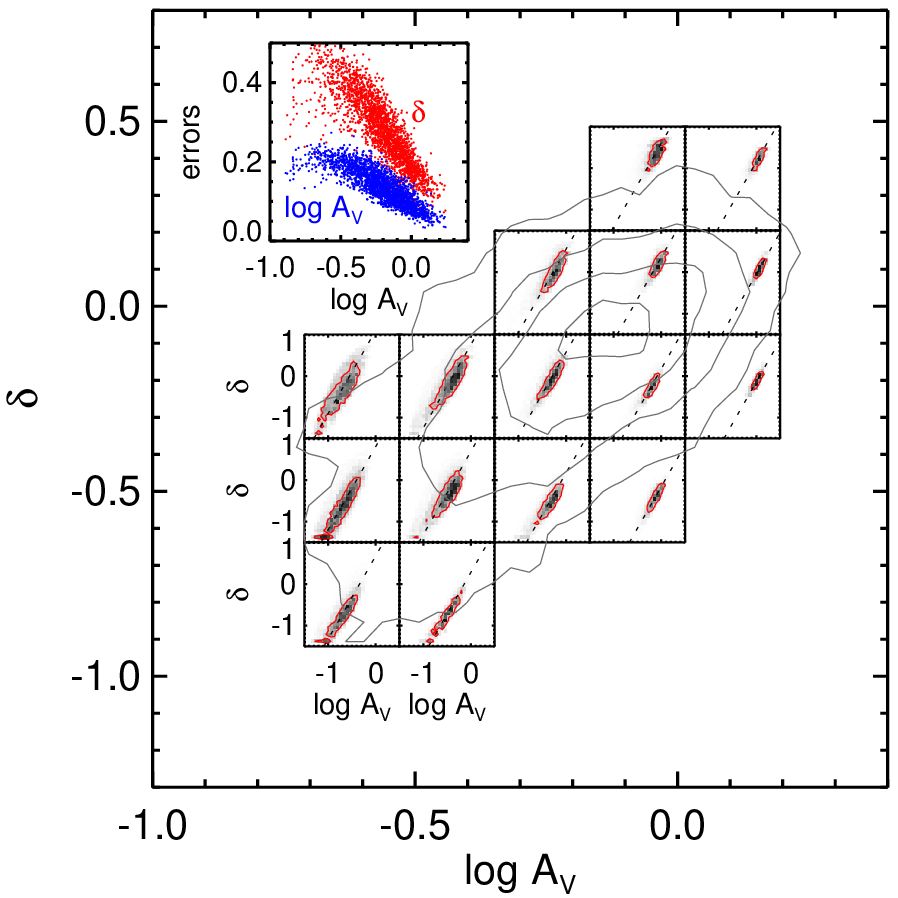}
\includegraphics[width=0.9\columnwidth]{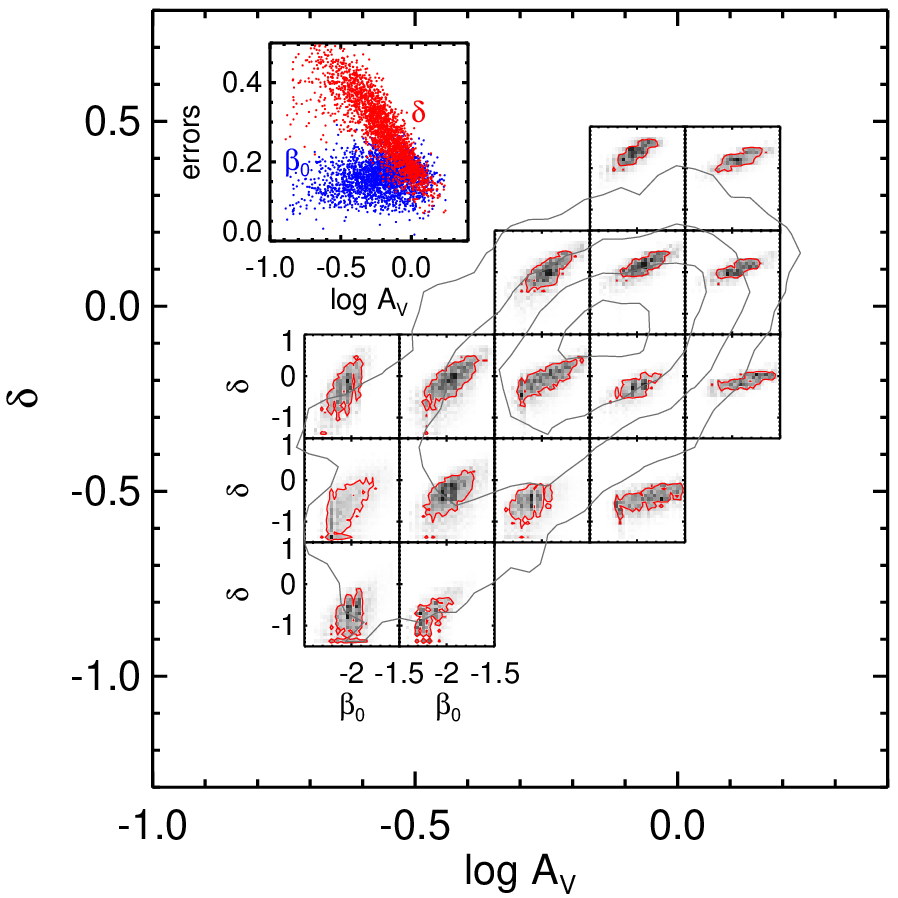}
\caption{Left: Background contours show the distribution of sample galaxies in the $\av$--$\delta$ plane. The inner grid panels show the PDFs between $\av$ and $\delta$ for galaxies at that location in the $\av$--$\delta$ plane. In each inner grid panel, red contour marks the 1\,$\sigma$-level of PDF, and the dotted line has a slope of 2 and passes through the median value. The small panel in the top-left corner shows the errors of $\delta$ (red) and $\av$ (blue) as a function of $\av$ for our sample of 2\,291 SFGs. Right: Similar to the left plot but showing  PDFs between $\beta_0$ and $\delta$. The top-left small panel shows the errors of $\delta$ (red) and $\beta_0$ (blue) as a function of $\av$ for our sample galaxies.}
\label{fig_av_delta_bayes_matrix}
\end{figure*}

\subsection{The degeneracies of $\beta_0$, $\delta$, $\av$, and $\afuv$ in SED fitting}\label{sec4.1}
The Bayesian approach can be used to examine the robustness of the output parameters with the  probability distribution function (PDF) \citep{Han2014,Sharma2017,Boquien2019,Yuan2019}. Degeneracies between parameters can be seen by correlations in posterior probability distributions \citep{Leja2017,Hagen2017,Han2019,Doore2021}. Figure~\ref{fig_bayes_matrix} shows the probability distribution for galaxy parameters $\beta_0$, $\delta$, $\av$, and $\afuv$, for a typical galaxy with the best-fitting galaxy parameters and errors representative among the sample.

From panel (g) of Figure~\ref{fig_bayes_matrix} one can see that $\delta$ is degenerate with the intrinsic UV slope $\beta_0$ in the sense that a lower $\beta_0$ (bluer in the UV) is coupled with a smaller $\delta$ (steeper attenuation curve). The Spearman's rank correlation coefficient $\rho_{\rm s}$ between the two parameters is $\sim$0.66. The SED of a young stellar population attenuated by a given dust attenuation curve can be replaced with the SED of an older stellar population attenuated by a flatter attenuation curve. This is the well-known degeneracy between stellar population age and the steepness of the dust attenuation curve \citep{Hagen2017,Calzetti2021}. In the framework of SED fitting based on the energy balance approach, the IR luminosity is used to constrain the integrated energy that is absorbed by dust; but how the energy  is absorbed across wavelength, i.e. the attenuation curve, is not constrained. Therefore the traditional method of SED fitting is not able to break this $\beta_0$--$\delta$ degeneracy effectively. Both $\beta_0$ and $\delta$ are poorly constrained in our SED fitting with CIGALE. 

Panel (h) of Figure~\ref{fig_bayes_matrix} shows a strong degeneracy between $\delta$ and $\av$ ($\rho_{\rm s}=0.84$). The higher $\av$, the flatter the attenuation curve. A similar trend is also presented in \citet{Hagen2017} and \citetalias{Salim2020}. This degeneracy is probably responsible for the well-established relation between $\av$ and the attenuation curve slope $\delta$ \citep{Salmon2016, Salim2018, Salim2020, Battisti2020}. We find that $\av$ is weakly degenerate with $\beta_0$ ($\rho_{\rm s}=0.36$), in the sense that a higher $\av$ is seen at a higher  $\beta_0$ (redder in the UV). Combined together, the three parameters $\av$, $\delta$, and $\beta_0$ seem to be degenerate with each other. This can be easily verified with the relation between two parameters by fixing the third one. As shown in the top-right panels of Figure~\ref{fig_bayes_matrix}, when we fix one parameter of the three, the degeneracy between the other two is to some extend compressed. For example [the panel (k)], the 1\,$\sigma$ dynamical range of $\beta_0$ ($\log\av$) decreases from  $\sim$0.5 ($\sim$0.5) to $\sim$0.3 ($\sim$0.2) if we fix $\delta=-$0.2. These results support that the three parameters $\av$, $\delta$, and $\beta_0$ are degenerate with each other. This is to say that a redder intrinsic SED, a higher $A_{\rm V}$ or a lower $\delta$ (steeper attenuation curve) in SED fitting may end up with similar model SEDs matching an observed SED. We refer it to as the $\av$--$\delta$--$\beta_0$ degeneracy.

\begin{figure}%[htb]
\centering
\includegraphics[width=0.9\columnwidth]{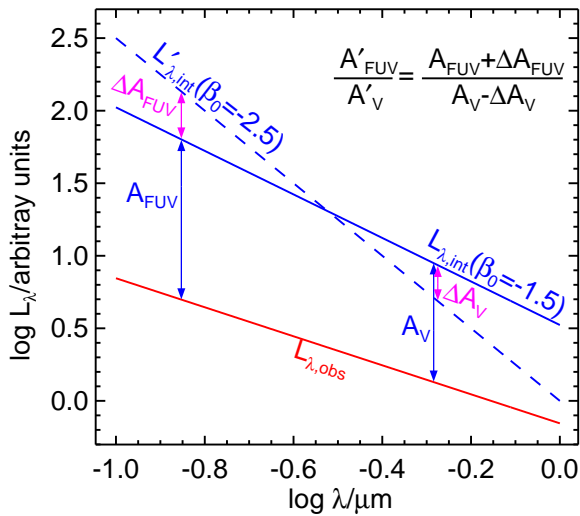}
\caption{Schematic diagram of the influence of $\beta_0$ fluctuation on  attenuation curve slope in the SED fitting based on the energy balance principle. Two intrinsic SEDs have the same bolometric luminosity but have different UV slopes ($\beta_0=-2.5$ and $-$1.5). The two UV slopes are roughly chosen as the dynamic range boundaries for model SEDs, and $\Delta\afuv$ and $\Delta\av$ represent the maximum variation of dust attenuation caused by $\beta_0$ fluctuation.  By definition, the amount of attenuation ($\afuv$, $\av$, $\Delta\afuv$, and $\Delta\av$) is proportional to the length of respective arrows given in the plot, i.e. $A_{\rm \lambda}=-2.5\log(L_{\rm \lambda,obs}/L_{\rm \lambda,int})\propto\log L_{\rm \lambda,int}-\log L_{\rm \lambda,obs}$. The relationship between the two attenuation curve slopes (parameterized by $\afuv/\av$) is also given. }
\label{fig_schematic}
\end{figure}

Moreover, we notice that FUV attenuation ($\afuv$) does not feature a strong degeneracy with $\delta$ ($\rho_s=-0.13$) as shown in panel (i). The ratio of $\afuv$ and $\av$ ($\afuv/\av$) represents the attenuation curve slope. Figure~\ref{fig_bayes_matrix} shows that the presence of strong $\av$--$\delta$ but no $\afuv$--$\delta$ degeneracy indicates that the variation of dust attenuation curve slope in SED fitting is mostly driven by  the change of $\av$ instead of $\afuv$. This is not surprising since the energy absorbed by dust comes mostly from the UV rather than the optical \citep{Cortese2008, Kennicutt2012}. If there is a fluctuation in $\delta$ in the SED fitting (coming from the $\beta_0$--$\delta$ degeneracy),  a consequent change in $\av$ is more preferred than in $\afuv$, because changing $\av$ have less effect on energy balance than chaning $\afuv$. As a consequence, the typical scatter of $\log\av$ is systematically larger than that of $\log\afuv$ (0.16 vs 0.08). The extra contribution comes from the scatter of $\delta$. The fluctuations in $\log\av$ and $\delta$ are highly degenerate and can bias the $\av$--$\delta$ relation derived from SED fitting.

To address these fitting degeneracies more clearly, we show the PDFs between $\av$ and $\delta$  across the $\av$--$\delta$ plane of our sample of 2\,291 galaxies (the background contour) in Figure~\ref{fig_av_delta_bayes_matrix}. We find that the $\av$--$\delta$ degeneracy is not monochromatic across the $\av$--$\delta$ plane of our sample galaxies. The scope of degeneracy decreases with $\av$, and at a fixed $\av$, it decreases mildly with $\delta$. Despite the amplitude changing dramatically across the $\av$--$\delta$ plane, the degeneracy fluctuations in $\log\av$ and $\delta$ change in a lock step  (with a slope of $\sim$2). The top-left panel shows the dependence of the uncertainties of $\delta$ and $\log\av$ on $\log\av$ for our sample galaxies. The uncertainties of both $\delta$ and $\log\av$ decrease with $\log\av$. The uncertainty in $\delta$ is about a factor of 2 times that in $\log\av$. 

It can be understood that the measurement uncertainty in $\delta$ exhibits a dependence on $\log\av$. Figure~\ref{fig_schematic} illustrates how a change in $\beta_0$ affects the determination of the attenuation curve slope $\delta$ in SED fitting. As discussed above, the uncertainties in $\delta$ in part originate from the variation in $\beta_0$. There are two ways to result in a smaller uncertainty for the attenuation curve slope $\delta$. One is to reduce the dynamical range of $\beta_0$ for model SEDs used in the SED fitting. We will show in Section~\ref{sec5} that a constant starburst setting with a narrow range of $\beta_0$ leaves on average a smaller uncertainty in $\delta$ [see also in the panel (m) of Figure~\ref{fig_bayes_matrix}]. Here for a given fitting configuration, the dynamical range of $\beta_0$ is fixed, and the uncertainties in $\beta_0$ is more or less constant (see in the right panel of Figure~\ref{fig_av_delta_bayes_matrix}). The second way is to increase the global dust attenuation (both $\afuv$ and $\av$). As shown in Figure~\ref{fig_schematic}, for a certain SED fitting, the maximum variation of dust attenuation (e.g. $\Delta\afuv$ and $\Delta\av$) caused by the change in $\beta_0$ is strictly limited. With the increase in global dust attenuation, both the $\Delta\afuv$ and $\Delta\av$ become less significant relative to the large $\afuv$ and $\av$. As a consequence, the attenuation curve slope $\delta$ will be decreasingly affected by the change in $\beta_0$. It is clear that the $\beta_0$--$\delta$ degeneracy is almost gone  at the high end of $\av$, as shown in Figure~\ref{fig_av_delta_bayes_matrix} (the right panel). Although the scatter of $\beta_0$ remains large at high $\av$, the uncertainty in $\delta$ drop significantly. This effect of `decreasing influence of $\beta_0$ on $\delta$ at higher $\av$' is also the key to understanding the different fitting results from our two fit configurations. We will come back to this in Section~\ref{sec5.2}. 

\begin{figure*}%[htb]
\centering
\includegraphics[width=0.9\textwidth]{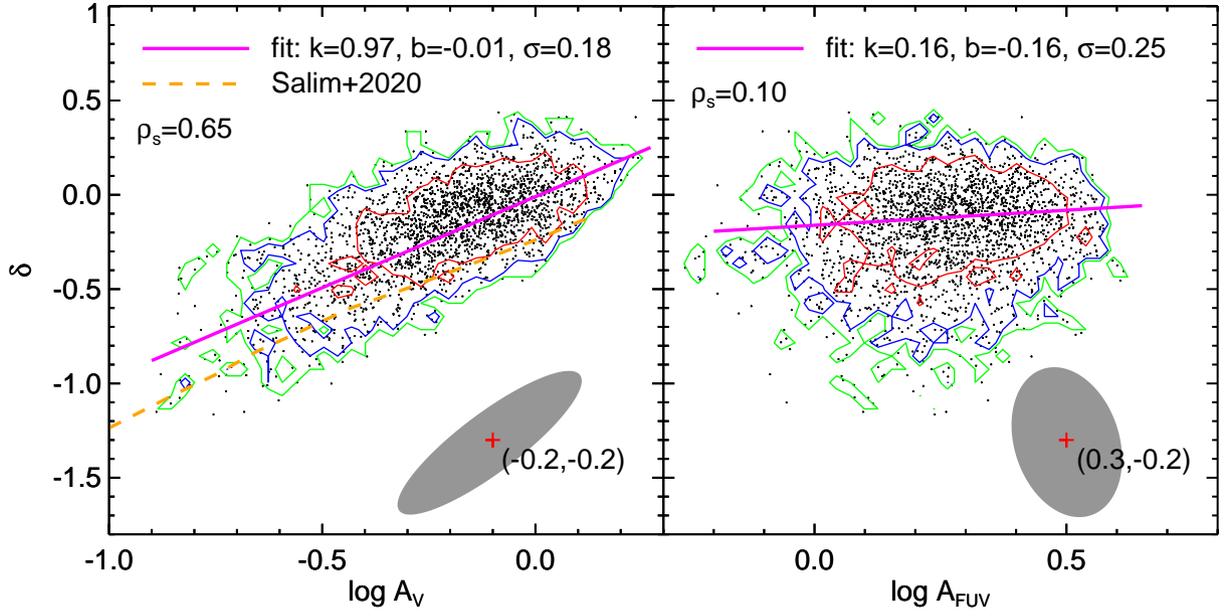}
 \caption{Left: Attenuation curve slope $\delta$ as a function of $A_{\rm V}$ on the basis of best fits with CIGALE under the fiducial declining starburst setting for our sample of 2\,291 local SFGs. The red, blue, and green contours enclose 68, 95, and 99\,per\,cent of sample galaxies, i.e. 1, 2, and 3\,$\sigma$, respectively. The magenta solid line is the relation best fitting the data points. The best-fitting parameters ($\rm Y=k\times X+b$) and dispersion ($\sigma$) are also presented. The orange dashed line represents the median relation given in \citetalias{Salim2020}. The Spearman's rank correlation coefficient ($\rho_s$) between $\av$ and $\delta$ is also labelled. The error ellipse at the bottom-right corner represents the 1\,$\sigma$ degeneracy error of the typical galaxy taken from  Figure~\ref{fig_bayes_matrix}. The $\av$ and $\delta$ of this typical galaxy are marked. Right: Similar to the left plot but showing the relation between $\delta$ and $\afuv$. }
\label{fig_ax_delta}
\end{figure*}

\begin{figure*}%[htb]
\centering
\includegraphics[width=0.9\textwidth]{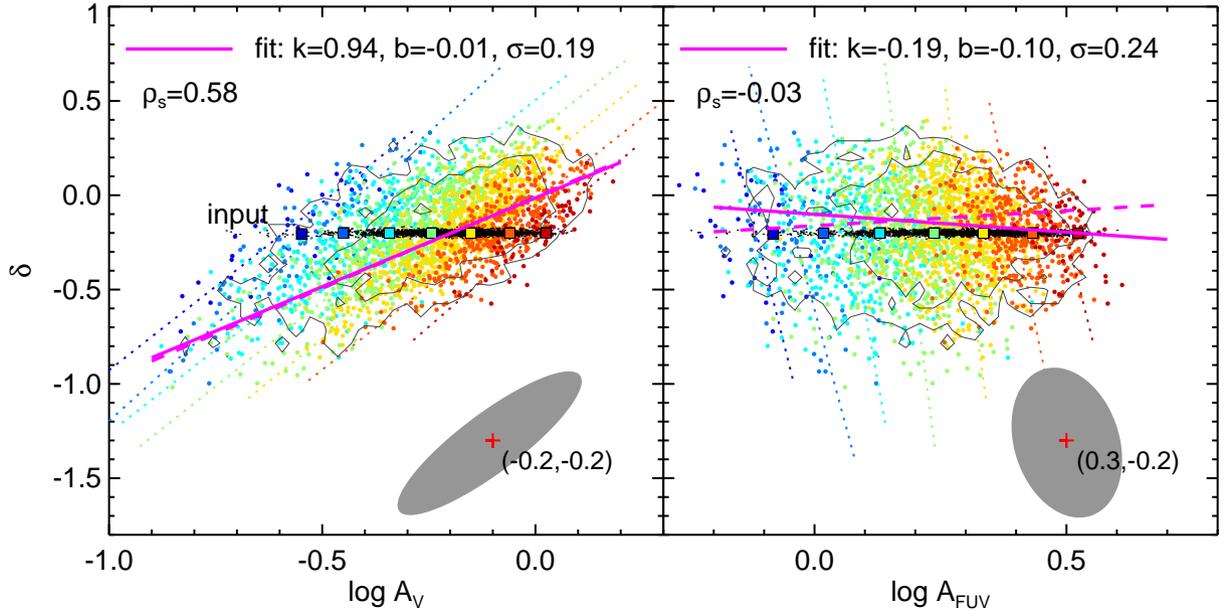}
\caption{Left: The best-fitting $\delta$ as a function of $\av$ for the simulated galaxies. The blue-to-red colour indicates increasing values of the input $\av$. The two contours enclose 68 and 95\,per cent data points. The black points are the input flat $\av$--$\delta$ relation with a fixed $\delta$ of $-0.2$. These data are slightly scattered in $\delta$ for demonstration. The coloured squares are the median of black points divided into different input $\av$ bins. The dotted lines are the best-fitting relations of simulated galaxies in different input $\av$ bins. The magenta solid line represents the relation best-fitting the sample. The best-fitting parameters and associated scatter are given in the legend. The magenta dashed line is the best-fitting relationship taken from Figure ~\ref{fig_ax_delta}. The Spearman's rank correlation coefficient ($\rho_s$) between $\av$ and $\delta$ is also labelled. The error ellipse in bottom-right represents the typical degeneracy error.  Right: Similar to the left but showing the relation between $\delta$ and $\afuv$.}
\label{fig_ax_delta_mock}
\end{figure*}

\subsection{The correlation between $\delta$ and $\av$ estimated with CIGALE}\label{sec4.2}

Figure~\ref{fig_ax_delta} shows our results from the CIGALE SED fitting of the observed SEDs for our sample of 2\,291 local SFGs.  The relation between attenuation curve slope $\delta$ and dust attenuation $\av$ is clearly seen. At increasing $\av$, the dust attenuation curve becomes flatter with increasing $\delta$. We compare our $\av$--$\delta$ relation to that given in \citetalias{Salim2020}.\footnote{\citetalias{Salim2020} used the UV-optical slope, defined as $S=A_{\rm 1500}/A_{\rm V}$, to parameterize the attenuation curve slope. The conversion follows $\delta=0.71-1.91\times\log S$ for our sample SFGs.} We find that our $\av$--$\delta$ relation is consistent with  \citetalias{Salim2020}'s but systematically higher in $\delta$ (shallower attenuation curve). The discrepancy is attributed to the adoption of a constant starburst for model SEDs in \citetalias{Salim2020} (see in Section~\ref{sec5}).

We show the typical degeneracy error in the form of an error ellipse for comparison. The error is defined as the 1\,$\sigma$ posterior distribution between $\delta$ and $\av$ (see in Figure~\ref{fig_bayes_matrix}).   It is clear that the size of the error ellipse between $\delta$ and $\av$ appears comparable to the distribution of our sample galaxies (the 1\,$\sigma$ red contour). We remind that the global dispersion of $\delta$ ($\sim$0.25) is slightly lower than the median uncertainty of $\delta$ ($\sim$0.28), indicating that the scatter in $\delta$ for our sample SFGs mainly come from the fitting uncertainties. The 1\,$\sigma$ dispersion around the best-fitting relation is $\sim$0.18, which is significantly smaller than the uncertainty of $\delta$. Generally speaking, the dispersion around the best-fitting relation is unlikely lower than the fitting uncertainty. However, this statement only holds when the two errors are independent. If the errors of two variables are correlated with each other, like our $\av$ and $\delta$, it will strengthen the correlation and leave the dispersion smaller than the measurement errors. 

On the other hand, the local SFGs in our sample show no correlation between $\afuv$ and $\delta$, as shown in the right plot of Figure~\ref{fig_ax_delta}. The independent error ellipse further confirms the robustness of this flat relation. Considering the large scatter in $\delta$, a flat $\afuv$--$\delta$ relation does not conflict with a positively-correlated $\av$--$\delta$ relation. We notice that the $\av$--$\delta$ relation is significantly biased by the fitting degeneracy (even comparable). It gives rise to a possibility that the `true' $\delta$ does not correlate with either $\av$ or $\afuv$ \citep[$\sim$ dust column density;][]{Butler2021}, while the measured $\av$--$\delta$ relation comes from the fitting degeneracy.

\subsection{A simulation test with a flat $\av$--$\delta$ relation}\label{sec4.3}

We conduct a simulation test to verify the possibility of lacking intrinsic dependence of attenuation curve slope on $\av$. We carry out the simulation by setting the `true' attenuation slope unchanged with $\av$, and testing if the fitting degeneracies produce a similar $\av$--$\delta$ relation? To do so, we firstly create a set of mock galaxy SEDs satisfying a flat $\av$--$\delta$ relation, i.e. a fixed $\delta$ over a range of $\av$, and then perform the same SED fitting to the mock data. 

We generate mock galaxy SEDs as follows. From the best-fitting results of Figure~\ref{fig_ax_delta}, we use the recovered intrinsic SEDs of our sample galaxies and attenuate them by the corresponding $\afuv$ with the dust attenuation curve of fixed $\delta=-0.2$. The best-fitting $\afuv$ is adopted since it better traces dust attenuation than $\av$ (see Figure~\ref{fig_bayes_matrix}). Taking into account the fitting uncertainties, here the dynamical range of input $\afuv$ slightly shrinks by $\sim$15\,per cent. The value of $\delta=-0.2$ roughly corresponds to the median value of our sample SFGs. We calculate fluxes in all bands and assign them errors the same as the measurement errors relative to the observed fluxes. As mentioned in Section~\ref{sec3.1}, about 10\,per cent uncertainties are added to the band fluxes in CIGALE, to account for the uncertainties from the models themselves. To be consistent, the additional 10\,per cent model errors are also included in generating mock SEDs. The attenuated fluxes are added with deviations randomly given by a normal distribution with the assigned errors as the dispersion. IR luminosity is calculated by integrating the total energy attenuated by dust (also adding errors). After that, we obtain simulated galaxy SEDs satisfying a known (flat) $\av$--$\delta$ relation. The flux distributions in all bands we examined are similar between our mock SEDs and the observed ones. We run CIGALE to repeat the same SED fitting (with $\delta$ as a free parameter) to our mock galaxy SEDs, and determine the best-fitting attenuation parameters.

Figure~\ref{fig_ax_delta_mock} shows the derived $\delta$ as a function of $\av$ from our fitting of the mock galaxy SEDs. We find the distribution of best-fitting $\delta$ appears similar to that in Figure~\ref{fig_ax_delta} even though the input $\delta$ is fixed to $-0.2$ (the black points). This indicates that the variations in $\delta$ for our sample SFGs mainly come from the fitting uncertainties. Compared with the input values, the output $\delta$ and $\av$ deviate following the degeneracy error with a slope of $\sim$2. The deviations are larger at lower input $\av$ (with bluer colour), consistent with the increasing degeneracy error at decreasing $\av$ shown in Figure~\ref{fig_av_delta_bayes_matrix}. The degeneracy error alters the input flat relation and forms an $\av$--$\delta$ relation similar to the $\av$--$\delta$ relation reported in the literature. The $\av$--$\delta$ relation given in Figure~\ref{fig_ax_delta} is presented by the magenta dashed line for comparison. We can see that the mock-based relation has a slope of 0.94 in comparison with the slope of 0.97 for the dashed line. Not only the best-fitting slopes agree but also the scatter of data points mirror each other: 0.19 and 0.18 for mock- and observation-based relation, respectively. The fitting degeneracies seem to be a dominant driver of the established $\av$--$\delta$ relation. In addition, the right panel of Figure~\ref{fig_ax_delta_mock} shows the independent error in $\afuv$ increases the scatter in $\delta$, but does not alter the input flat $\afuv$--$\delta$ relation significantly. Our simulation test confirms that the degeneracy error can significantly bias a flat $\av$--$\delta$ relation and result in a  $\av$--$\delta$ relation similar to what we often see in the literature. We thus conclude that the strong $\av$--$\delta$ correlation derived from SED fitting is a false relation governed by fitting degeneracies.

\section{Fitting with model SEDs of constant starburst SFHs}\label{sec5}

\begin{figure*}%[htb]
\centering
\includegraphics[width=0.8\textwidth]{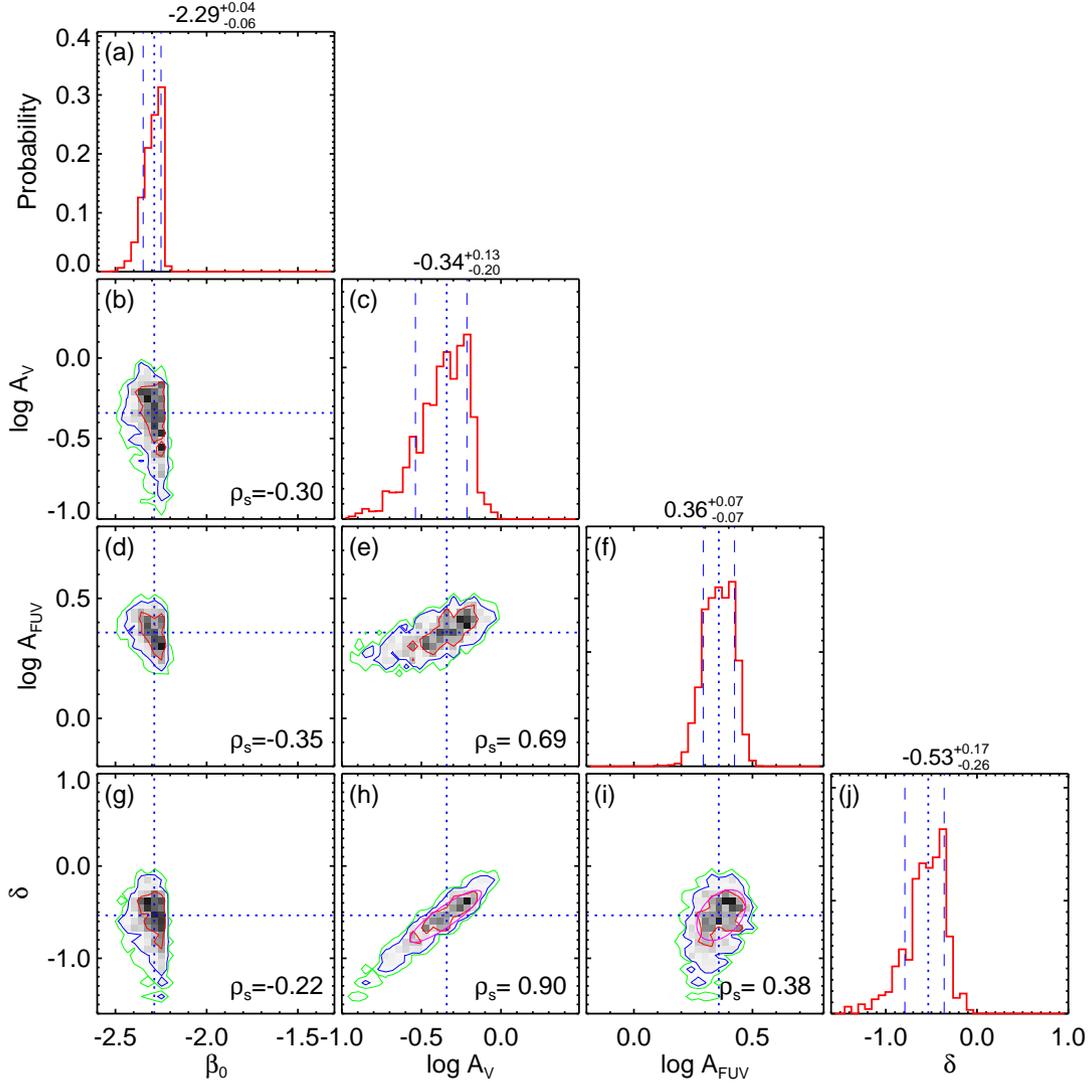}
\caption{Similar to Figure~\ref{fig_bayes_matrix} but showing the results derived using model SEDs with the constant starburst SFHs.}
\label{fig_bayes_matrix_const}
\end{figure*}

\begin{table*}
 \centering
	\caption{Comparison of the median best-fitting parameters and $\chi_{\rm r}^2$ between the declining starburst and constant starburst settings for SED fitting of our local sample. }  \label{tab2}
 \begin{tabular}{cccccc}
  \hline
      	           &$\beta_0$           &$\delta$            &$\log$$\av$          &$\log$$\afuv$      &$\chi_{\rm r}^2$     \\
  \hline
fiducial declining starburst   &-2.00$\pm$0.16      &-0.14$\pm$0.28      &-0.19$\pm$0.14     &0.26$\pm$0.08     &0.18  \\
constant starburst    &-2.33$\pm$0.05      &-0.54$\pm$0.22      &-0.34$\pm$0.13     &0.32$\pm$0.06     &0.28  \\
\hline
%\multicolumn{6}{c}{$^a$The median of our local 2\,291 SFGs, slightly different from the value of typical galaxy shown in Figure~\ref{fig_bayes_matrix} and Figure~\ref{fig_bayes_matrix_const}.} \\    
 \end{tabular}
\end{table*}

\begin{figure*}%[htb]
\centering
\includegraphics[width=0.9\textwidth]{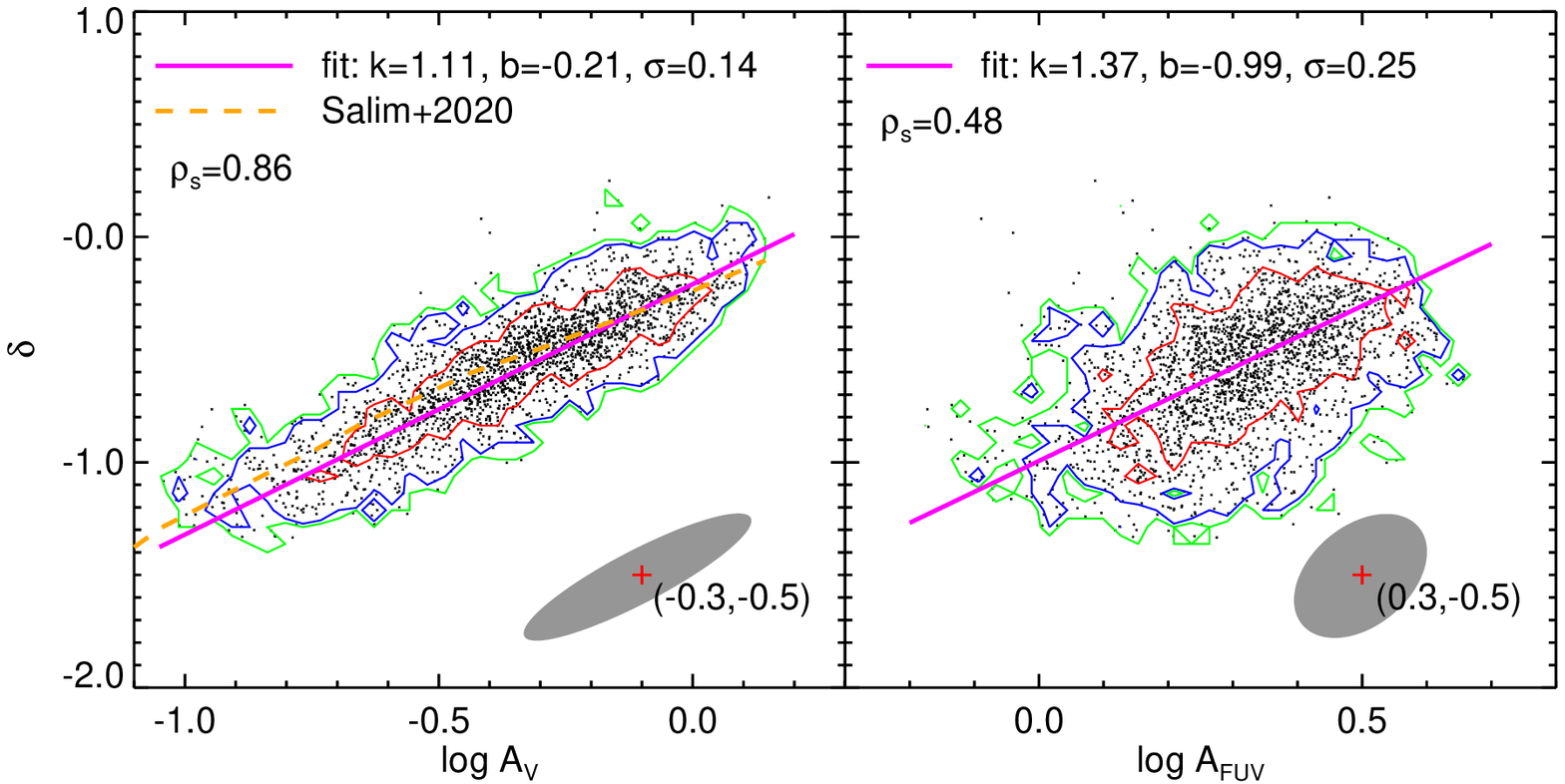}
\caption{Similar to Figure~\ref{fig_ax_delta} but showing the results from the fitting with model SEDs of constant starburst SFHs.}
\label{fig_ax_delta_const}
\end{figure*}

Our finding that the $\av$--$\delta$ relation is dominated by fitting degeneracies disagrees with the result from \citetalias{Salim2020} that the size of degeneracy error is significantly smaller compared to the tight global correlation.  As we pointed out, the use of model SEDs made with constant starburst SFHs in their SED fitting is mainly responsible for this difference. We decide to conduct SED fitting in the same way and make a quantitative comparison between the fitting results from the constant starburst and declining starburst settings.

\subsection{The outputs from the constant starburst fitting}\label{sec5.1}

Following Section~\ref{sec4.1}, we show the degeneracies between galaxy attenuation parameters for the same typical galaxy in Figure~\ref{fig_bayes_matrix_const}. We stress that $\beta_0$ of the model SEDs from the constant starburst SFHs is distributed in a narrow range around $\beta_0=-2.3$ and the degeneracy of $\beta_0$ with $\delta$ (as well as $\av$) is largely compressed. The narrow range of $\beta_0$ does not mean that $\beta_0$ is well determined. Instead, it is decided by the chosen model SEDs  (see Figure~\ref{fig_ssfr_beta0}). In other words, $\beta_0$ is not a fully free parameter in the SED fitting with the constant starburst setting. As a consequence, the median error of $\beta_0$ (for our sample), calculated as the standard deviation of the PDF, decreases significantly from $\sim$0.16 for the declining starburst setting to $\sim$0.05 (see Table~\ref{tab2}). As discussed in Section~\ref{sec4.1}, a `well' constrained $\beta_0$ will subsequently lead to a smaller error in $\delta$. The median error of $\delta$ decreases from $\sim$0.28 to $\sim$0.22. Although $\delta$ is degenerate with $\av$, the degeneracy error is relatively smaller. The median error of $\delta$ is moderately larger than the median value of $\sim$0.17 given in \citetalias{Salim2020} using a similar constant starburst setting.\footnote{\citetalias{Salim2020} conducted the analysis with the same GSWLC-D sample used in \citet{Salim2019}. More details can be found in the latter work as well as in \citet{Salim2018}. } This noticeable difference might be partially due to the dynamical range of [$-$1.4, 0.6] for $\delta$ in our SED fitting slightly larger than the range of [$-$1.2, 0.4] given in \citetalias{Salim2020}. We verify that a smaller typical error of $\sim$0.20 will be obtained if a consistent $\delta$ range is set in our SED fitting.

On the other hand, the limitation of $\beta_0<-2.2$ in the constant starburst setting subsequently biases the derived attenuation curves. Table~\ref{tab2} summarizes the median parameters (for our local 2\,291 SFGs) and corresponding errors estimated from two sets of SED fitting. One can see that $\beta_0$ from the constant starburst fitting is much smaller (bluer intrinsic UV colour) than that of the declining starburst fitting. As a consequence, it returns a steeper attenuation curve according to the $\beta_0$--$\delta$ degeneracy (see Figure~\ref{fig_bayes_matrix}). The median of $-$0.54 for $\delta$ from the constant starburst fitting is slightly lower that the median of $-0.42$ presented in \citetalias{Salim2020}. Still, the median attenuation curve is systematically steeper than that from the declining starburst fitting (median is $\delta=-0.14$). Generally speaking, a steeper attenuation curve can be reflected by either an increasing $\afuv$ or decreasing $\av$. From the declining starburst fitting to the constant starburst fitting, the median best-fitting $\log\afuv$ increases by only 0.06\,dex while $\log\av$ decreases by 0.15\,dex. This is consistent with the expectation that the fluctuation in $\delta$ (error-driven) in the energy-balance fitting is more linked with $\av$ than $\afuv$ in Section~\ref{sec4.1}.  

We notice that the declining starburst fit yields an smaller reduced chi-square $\chi_{\rm r}^2$ than that of a constant starburst fit. Moreover, the two $\chi_{\rm r}^2$ are smaller than unity, indicating either an over-fitting or an overestimate of errors. The latter seems reasonable since, by default, additional 10\,per cent model errors are added to the `input' photometry errors (i.e., overestimate of errors). More importantly, we note that the $\chi_{\rm r}^2$ presented here is defined as $\chi^2/(N-1)$, where N is the number of data points. This differs from the `true' reduced $\chi^2$ which is defined as $\chi^2/N_{\rm dof}$, where $N_{\rm dof}$ is the number of degrees of freedom \citep{Andrae2010}. The $N_{\rm dof}$ can be estimated for linear models as $N_{\rm dof}=N-K$, where $K$ is the number of free parameters. For the nonlinear models (like our SED-fitting), it is questionable whether it can be accurately calculated \citep[][]{Andrae2010,Malek2018}. Given that K always greater than 1, the $\chi_{\rm r}^2$ presented here [$\chi^2/(N-1)$] should be always smaller than the `true' value of $\chi^2/(N-K)$. We thus do not treat the small $\chi_{\rm r}^2$ outputted by CIGALE as a sign of over-fitting \citep[see also][]{Malek2018,Nersesian2019,Boquien2019,Ren2022}.

Figure~\ref{fig_ax_delta_const} shows the $\av$--$\delta$ relation from the constant starburst fitting, being in good agreement with the relation from \citetalias{Salim2020}. Again we emphasize that the model SEDs with constant starburst SFHs are adopted in both of the two. We point out the degeneracy error is relatively smaller compared to the global $\av$--$\delta$ relation, consistent with \citetalias{Salim2020}. Compared with the results by the declining starburst fitting, the relation here is tighter and has a smaller dispersion ($\sigma$ decreases from 0.18 to 0.14). Moreover, the right panel shows the $\delta$ as a function of $\afuv$. For the declining starburst fitting, the relation is flat, and for a constant starburst fitting, $\delta$ moderately increases with $\afuv$. An additional dependence of $\delta$ on $\afuv$ (or global dust attenuation) appears when the constant starburst fitting is applied. 

\begin{figure}%[htb]
\centering
\includegraphics[width=0.45\textwidth]{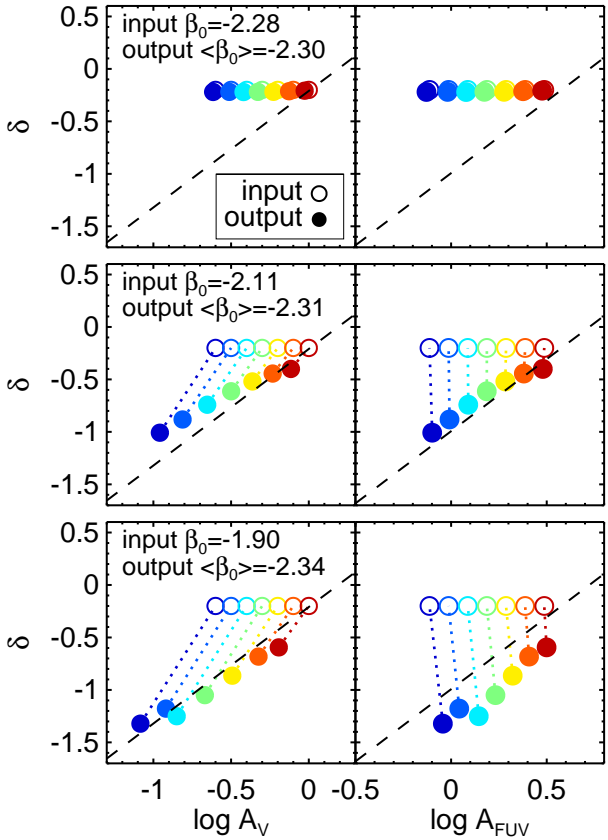}
\caption{Attenuation curve slope $\delta$ as a function of $\av$ (left) and $\afuv$ (right) estimated from the constant starburst fitting to the simulated galaxy SEDs with  a fixed $\beta_0 = -2.28$ (top), $-2.11$ (middle), and $-$1.90 (bottom). The open and filled circles represent the input and output attenuation parameters, respectively, connected by dotted lines. The colour-coding reflects input $\av$ (or $\afuv$). The dashed lines in these panels refer to the best-fitting relations taken from Figure~\ref{fig_ax_delta_const}. The average of output $\beta_0$ is presented in each left panel.}
\label{fig_ax_delta_mock_grid}
\end{figure}

\begin{figure*}%[htb]
\centering
\includegraphics[width=0.9\textwidth]{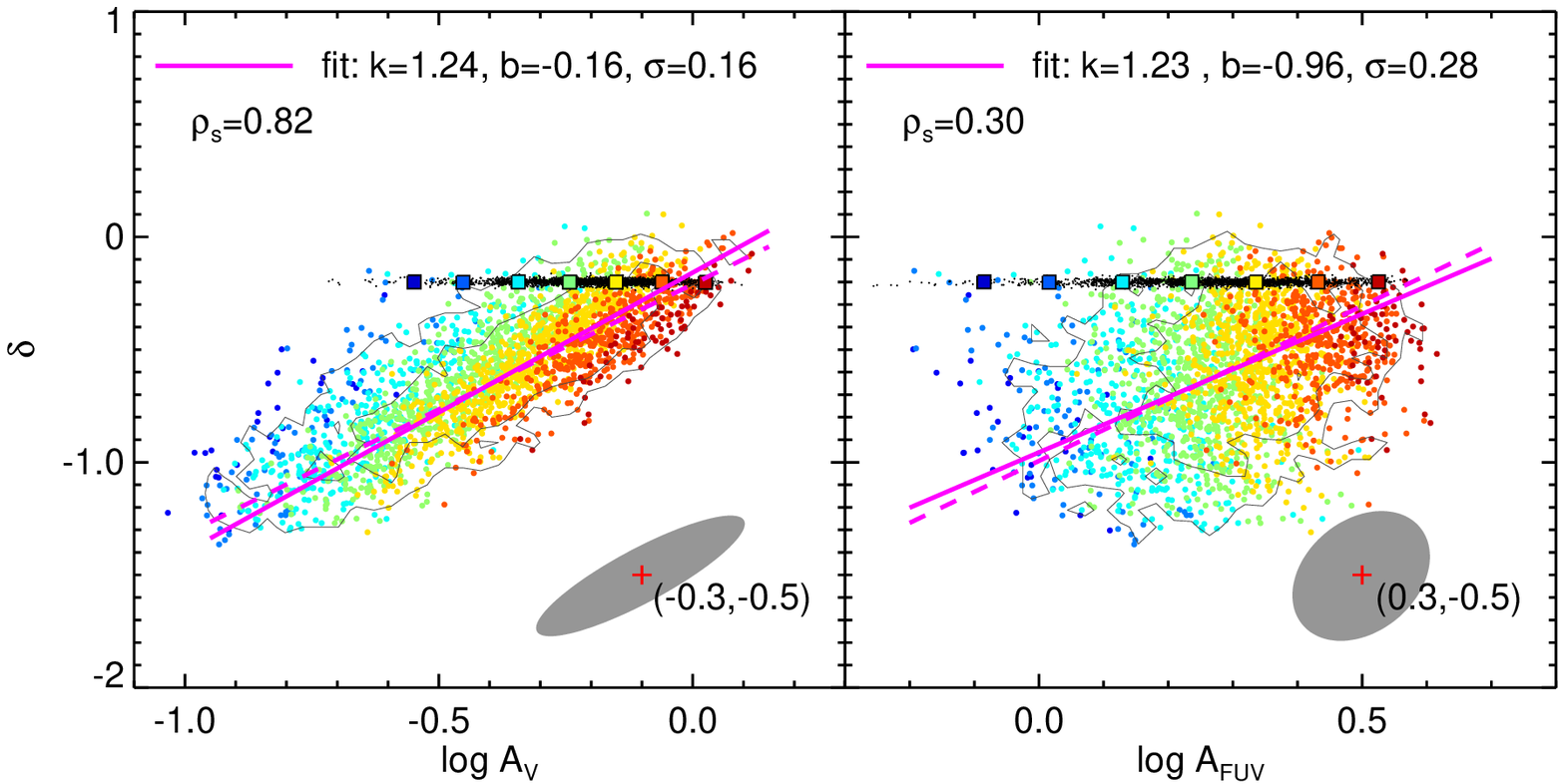}
\caption{Similar to Figure \ref{fig_ax_delta_mock}, but the SED fitting is done with the constant starburst setting. The magenta dashed lines refer to the best-fitting relations from Figure~\ref{fig_ax_delta_const}.}
\label{fig_ax_delta_const_mock}
\end{figure*}

\subsection{Understanding the SED fitting results with the declining starburst and constant starburst settings}\label{sec5.2}

We have demonstrated that the main difference between the declining starburst and constant starburst settings is the dynamic range of $\beta_0$. Unlike the widely distributed $\beta_0$ from $-2.5$ to $-1.5$ generated by the declining starburst setting, the constant starburst setting scans a very limited range $\beta_0=[-2.5,-2.2]$. The best-fitting $\beta_0$ from the declining starburst runs has a median of $\sim-$$-2$, which is much higher (redder in the UV) than the $\beta_0$ coverage in the constant starburst fitting. It is worth noting that local SFGs tend to have high $\beta_0$ (red UV colour) and  spread in a wide range (more will be discussed in Section~\ref{sec6.1}). Then the constant starburst fitting produces inappropriately lower $\beta_0$ (bluer UV colour). This underestimate of $\beta_0$ will be translated into a smaller $\delta$ in SED fitting according to the $\beta_0$--$\delta$ degeneracy. However, as mentioned in Section~\ref{sec4.1} (Figure~\ref{fig_schematic}), the change in $\delta$ caused by  the deviation of $\beta_0$ is dependent on the global dust attenuation. In the low dust attenuation regime, the constant starburst fitting (with model SEDs bluer in the UV) gives a smaller $\delta$ with large deviation; in the high dust attenuation regime (both $\av$ and $\afuv$), $\delta$ is no longer sensitive to the change in $\beta_0$ and has a small deviation. The higher the dust attenuation, the higher the $\delta$. Combined together, these biases and scatters caused by fitting degeneracies induce a positive relation between the attenuation curve slope $\delta$ and global dust attenuation (both $\av$ and $\afuv$). Moreover, if dust attenuation is extremely high, the selection bias in $\beta_0$ no longer affects the derived $\delta$; and the declining starburst and constant starburst two fittings will output similar attenuation curves. 

Aiming to further ascertain these effects, we perform SED fitting with the constant starburst setting to mock galaxy SEDs of different `true' $\beta_0$. The intrinsic galaxy SEDs come from the CIGALE SED libraries generated with the declining starburst setting. Considering that $\beta_0$, sSFR and metallicity of the model SEDs are correlated with each other (see Figure~\ref{fig_ssfr_beta0}),  we pick those with sSFR and metallicity to be representative of our sample SFGs, i.e, with $\log {\rm sSFR}\approx -9.6$\,yr$^{-1}$ and $Z=\rm Z_\odot$. Here we focus on three model SEDs with $\beta_0=-2.28$, $-$2.11, and $-$1.90.  Next step, we attenuate each model SED with the attenuation curve of a fixed slope $\delta=-0.2$ in combination with a set of $\av$ over $\log\av=[-0.6, 0]$. We derive the fluxes in all bands involved in our catalogue. IR luminosity is calculated by integrating the total energy attenuated by dust. To better illustrate the artefact of $\beta_0$ bias, we do not add any errors (and perturbation) to these band fluxes.  With these mock galaxy SEDs, we repeat SED fitting with the constant starburst setting and present the recovery of attenuation parameters in Figure~\ref{fig_ax_delta_mock_grid}. 

We remind that the recovered intrinsic UV slope from the constant starburst fitting remains steep ($\beta_0<-2.2$).  For the mock SED of  $\beta_0=-2.28$ (the top-left panel), the SED fitting can well recover the input $\beta_0$, $\delta$, and $\av$. For the mock SEDs of $\beta_0=-2.11$ and $-1.90$ (the middle-left and bottom-left panels), the recovered $\beta_0$ ($\sim$$-2.3$) deviates dramatically from the input value.  As a consequence, the recovered $\delta$ is increasingly smaller  at decreasing $\log\av$. The deviation becomes increasingly larger for higher $\beta_0$ due to the fitting degeneracies between $\beta_0$, $\delta$, and $\av$. Again, the underestimate of $\delta$ is dependent on $\av$ --- the deviation of  $\delta$ from the input value decreases at increasing $\av$.  A strong correlation can be seen between the recovered $\av$ and $\delta$ for the two model SEDs of input $\beta_0>-2.2$. Such correlation also holds for $\afuv$, as shown in the right panels of Figure~\ref{fig_ax_delta_mock_grid}. It becomes clear that the biases in recovering $\delta$ and $\av$ originate from the fitting degeneracies between $\beta_0$, $\delta$, and $\av$.

Interestingly, the bottom-left panel of Figure~\ref{fig_ax_delta_mock_grid} shows the recovered $\av$--$\delta$ relation (solid circles) following the $\av$--$\delta$ relation (the dashed line) for our sample of local SFGs presented in Figure~\ref{fig_ax_delta_const}. One question naturally arises --- \textit{do the biases driven by fitting degeneracies fully account for the formation of the observed $\av$--$\delta$ relation?}  We quantitatively evaluate the effects of these biases through SED fitting on the mock galaxy SEDs with a fixed $\delta$ of $-0.2$ presented in Section~\ref{sec4.3}. We analyse the systematic biases as examined before. We note that the mock galaxy SEDs span a wide range in $\beta_0$,  and thus allow to generate biases induced by the $\beta_0$-related degeneracies in the SED fitting with the constant starburst setting. We show the fitting results in Figure~\ref{fig_ax_delta_const_mock}.

One can see that with a fixed input $\delta$ for all mock galaxy SEDs, the fitting returns the recovered attenuation parameters forming an $\av$--$\delta$ relation (both slope and scatter) similar to that derived for our sample of local SFGs (Figure~\ref{fig_ax_delta_const}). We remind that Figure~\ref{fig_ax_delta_const_mock} is for mock galaxy SEDs attenuated by the same dust attenuation curve of $\delta=-0.2$, while Figure~\ref{fig_ax_delta_const} is for our sample of 2\,291 local SFGs.  The SED fitting  with the constant starburst setting more or less recovers the input $\delta$ at the high end of $\av$ but increasingly underestimates $\delta$ at decreasing $\av$. We emphasize that this correlation between the recovered $\av$ and $\delta$ in Figure~\ref{fig_ax_delta_const_mock} is completely attributed to the systematic biases induced by degeneracy errors between $\delta$, $\av$, and $\beta_0$ in the SED fitting with the constant starburst setting.

The right panel of Figure~\ref{fig_ax_delta_const_mock} shows that a correlation of $\delta$ with $\afuv$ is also recovered. This $\afuv$--$\delta$ correlation is mainly due to the bias of constant starburst setting that the intrinsic UV slope of all model SED templates is set to $\beta_0<-2.2$. When the model SED templates span over a wide range of $\beta_0$ as given by the declining starburst setting, the recovered $\delta$ no longer depends on $\afuv$ (the right panel of Figure~\ref{fig_ax_delta_mock}). Our simulation results explain why \citetalias{Salim2020} delivered a tight $\av$--$\delta$ relation with smaller degeneracy errors (under the constant starburst setting). We conclude that the degeneracies between dust attenuation curve slope $\delta$, dust attenuation $\av$, and the intrinsic UV slope of model galaxy SEDs $\beta_0$ in SED fitting cause systematic biases in deriving these quantities and result in false correlations between $\av$ ($\afuv$) and $\delta$.

\section{Discussion}\label{sec6}

\subsection{Distribution of the intrinsic UV slope among local SFGs} \label{sec6.1}

A well-designed declining starburst SFH is introduced in this work to generate model SEDs for fitting the observed galaxy SEDs, in comparison with the constant starburst SFH often adopted in previous studies \citep{Giovannoli2011, Buat2012, Salim2016, Malek2018, Salim2018, Salim2019, Salim2020}.  The model SED templates from the constant starburst SFHs have the intrinsic UV slope ($\beta_0$) in a limited range of $-2.5<\beta_0<-2.2$ (Figure~\ref{fig_ssfr_beta0}). This is because  the youngest stellar population is continuously added to the preexisting populations, keeping the intrinsic stellar UV colour blue. The use of such a set of model SEDs means that the target galaxies' intrinsic stellar SEDs are very blue in the UV with $-2.5<\beta_0<-2.2$ \citep[see also][]{Salim2019}. It is natural to ask how $\beta_0$ distributes among local SFGs?

When model SED templates have $\beta_0$ spanning over a reasonably wide range  (particularly $>-2.2$), like given in our declining starburst setting, the recovered $\beta_0$ through the SED fitting for our sample of 2\,291 local SFGs spreads from $-2.4$ to $-1.7$ (median is $\sim$$-2.0$). And the best-fitting $\chi_{\rm r}^2$ becomes systematically smaller, compared to the results from the constant starburst fitting.  We build new model SED templates by scanning $\tau_{\rm burst}$ from 100\,Myr to 10\,Gyr (i.e., scanning two values) and perform SED fitting for our sample SFGs to see which set of model SED templates best fit the observed data best. Our results show that about 83\,per\,cent of our sample SFGs are best fitted by the declining starburst model SEDs (i.e., $\tau_{\rm burst}=100$\,Myr). The best-fitting $\beta_0$ spreads from $-$2.5 to $-$1.7 (median is $-$2.04) and there are about 85\,per\,cent of galaxies have $\beta_0>-2.2$. These results indicate that a declining starburst SFH with red UV colour is more favoured by the local SFGs.  Indeed, a more complex SFH to generate model SEDs with red intrinsic UV colour of $\beta_0=-1.9$ is also suggested by \citet{Calzetti2021} for a local galaxy. They pointed out that the starburst regions usually have simple SFHs (e.g. a young instantaneous or constant starburst) and blue UV colours, while the entire galaxies consist of multiple generations of stellar populations (i.e., more complex SFH) usually have redder $\beta_0$. Similar results are also reported by \citet{Boquien2012}, who modelled galaxy SEDs with free-varied starburst to a sample of local SFGs and obtained the best-fitting $\beta_0$ spreading in $-2.2<\beta_0<-1.0$.  

It is not surprising that the intrinsic UV colour of local SFGs may be red and span a wide range. For instance, \citet{Dale2009} derived $\beta$ to be in [$-2.3$, $-0.6$]\footnote{The original UV slope in \citet{Dale2009} is given as $L_{\rm FUV}/L_{\rm NUV}$. We convert it into $\beta$ following the empirical relation given by \citet{Battisti2016}.} for metal-poor dwarf galaxies in the Local Volume when $\afuv$ is small (close to `zero'). Similarly, \citet{Battisti2016} obtained $\beta$ over [$-$2.1, $-$0.8] with a median of $-1.6$ at `zero' dust attenuation indicated by the Balmer decrement. One caveat is that the target galaxies in these studies are not representative for those in the regime of high dust attenuation. Nonetheless, we argue that local SFGs should have a rather red and large variation of $\beta_0$. 

On the other hand, a rather complex SFH with large variation of $\beta_0$ was also found in theoretical studies. For example, a theoretical investigation based on the IllustrisTNG simulations predicted a large spread for $\beta_0$ among local SFGs, giving $-2.3<\beta_0<-1.7$ with a median of $-$2.07 \citep{Schulz2020}. By analyzing a set of 51 hydrodynamical simulations of selected galaxies, \citet{Safarzadeh2017} gave a similar coverage of $\beta_0$ from $-2.1$ to $-1.3$ for isolated disc galaxies at $z=0$. In brief, local SFGs appear to exhibit a large scatter in $\beta_0$ (or the intrinsic UV colour).  

Back to SED fitting, an SFH having a significant fraction of intermediate-age stellar populations is the key to generate model SEDs with red UV colour \citep{Calzetti2021}. In our two-component SFH prescription, this requirement can be met by adding a declining starburst. If the starburst declines too fast, it fails to supply sufficient recently-formed stars, and the galaxy becomes old and has a lower sSFR; if a constant starburst is involved, the galaxy's UV colour remains blue ($\beta_0<-2.2$). Our declining starburst recipe sets the e-folding time to  100\,Myr and starburst fraction to [0.01, 0.5], being able to generate model SEDs with a reasonably wide coverage of $\beta_0$. This is important to reduce the systematic bias in SED fitting induced by the intrinsic UV colour. %overestimate of the intrinsic UV radiation.  

\subsection{Does the `true' attenuation curve slope correlate with dust column density?} \label{sec6.2} 

Our main goal is to address the effects of the degeneracies in SED fitting on the correlation between attenuation curve slope ($\delta$) and dust column density ($\sim\av$), which has been widely explored using a SED fitting technique in the literature \citep[e.g.][]{Arnouts2013, Kriek2013, Salmon2016, Hagen2017, Leja2017, Tress2018, Salim2018, Decleir2019, Battisti2020, Salim2020,Battisti2020}. We build the observed SEDs for a sample of local SFGs using high-quality multi-wavelength data from the FUV to the FIR and perform SED fitting with reasonable parameter settings. We find that fitting degeneracies induce systematic biases responsible for the correlation between the attenuation curve slope $\delta$ and $\av$. Our simulation tests  further confirm that this correlation is purely controlled by the degeneracy biases in the SED fitting (see Section~\ref{sec4}).

Our conclusion disagrees with the interpretation of the $\av$--$\delta$ relation in \citetalias{Salim2020}, which advised the degeneracy errors to be insignificant compared to the global correlation, and $\av$ as the dominant factor in regulating attenuation curve slope. We reproduced the $\av$--$\delta$ relation using our sample of local SFGs together with the constant starburst setting in SED fitting, following their settings \citep[see detail in][]{Salim2018, Salim2019}. Our simulation tests demonstrated that the degeneracies between $\av$, $\delta$, and $\beta_0$ in the SED fitting provoke systematic biases that give rise to a false $\av$--$\delta$ correlation; a further limitation on the intrinsic UV slope ($-2.5<\beta_0<-2.2$) for model SED templates (of constant starburst fitting) strengthens the $\av$--$\delta$ correlation, and leads to the shrinking of degeneracy errors and the emergence of a $\afuv$--$\delta$ correlation.   
 
In our tests, we started from a fixed attenuation curve slope $\delta$ for all mock galaxy SEDs and ended up with a reproduced $\av$--$\delta$ relation similar to that derived from a sample of local SFGs. The assumption of no correlation between $\delta$ and $\av$ (approximately dust column density) was taken for the mock SFGs. The test results do not exclude the possibility that $\delta$ might weakly depend on $\av$ for star-forming galaxies. When making the input mock galaxy SEDs shaped by an $\av$-dependent attenuation curve (either positive or negative), we still obtain an $\av$--$\delta$ relation having similar slopes as shown in Figure~\ref{fig_a1}. However, it can be distinguished by the dispersion. We find that the dispersion around the relations are 0.21, 0.19 and 0.18 for the mock galaxy SEDs with, satisfying negative, flat and positive input $\av$--$\delta$ relation, respectively. Compared to the dispersion of 0.18 given in Figure~\ref{fig_ax_delta}, a weak (slope of $\sim$0.2) or no `true' dependence of $\delta$ on $\av$ is favoured. 

On the other hand, the fitting errors only increase the scatter and do not significantly alter the input $\afuv$--$\delta$ relation (the right panels). Thus a rather flat `true' $\afuv$--$\delta$ relation is favoured. No intrinsic scatter of $\delta$ also means a flat `true' $\av$--$\delta$ relation (see Figure~\ref{fig_ax_delta_mock}). If the flat $\afuv$--$\delta$ relation has some intrinsic scatter in $\delta$, an increase in $\delta$ (by random scatter) towards a larger $\av$, i.e., a positive $\av$--$\delta$ dependence. Specifically, inputting an intrinsic scatter in $\delta$ with $\sigma=0.1$ and 0.2 will result in a positive $\av$--$\delta$ relation with a slope of 0.3 and 0.8, respectively. However, our mock tests have shown that the dispersion ($\sigma\sim0.25$) of best-fitting $\delta$ can be well recovered if a fixed $\delta$ is adopted (i.e., no intrinsic scatter). We inspect that a scattered $\delta$ with $\sigma=0.1$ (0.2) causes a dispersion of $\sigma=0.25$ (0.30) in output $\delta$. Therefore the intrinsic scatter of $\delta$ is expected to have $\sigma<0.1$, corresponding to an $\av$--$\delta$ relation with a positive slope of $<0.3$. These results suggest that the `true' $\av$--$\delta$ relation should be either flat or weakly positive. 

One may question if other methods than the energy-balance SED fitting could properly measure galaxy attenuation parameters that are free from the fitting degeneracies. One classical method for determining dust attenuation curve is to compare the attenuated SEDs with the reference `dust-free' SED (zero attenuation)  of a given type of galaxies \citep{Calzetti1994}. Applying this method to a sample of 5\,500 local SFGs, \citet{Battisti2017b} found the attenuation curve slope changes little with  either stellar mass or metallicity. Both quantities are expected to be good probes of dust attenuation or column density \citep{Garn2010, Xiao2012, Qin2019a, Bogdanoska2020, Shapley2021}. Similarly, \citet{Wild2011} examined dust attenuation curves using a ``pair-matching'' method. The galaxy pairs are selected with similar properties but have different dust attenuation. They found that galaxies with higher $M_\ast$ surface density ($\sim$ higher attenuation) and more face-on ($\sim$ smaller attenuation) tend to have steeper attenuation curves. \citet{Rezaee2021} applied a ``direct'' method developed by \citet{Reddy2020} to the local SDSS galaxies and found the slope of the (nebular) attenuation curve varies little with either the $M_\ast$ or metallicity.  All these empirical methods have certain shortcomings. Some need to assume that the dusty galaxies and less dusty galaxies have the same intrinsic SEDs. Some measure the attenuation curves of nebular lines, which may be inconsistent with those obtained through SED fitting (i.e. of stars). In any case, if a correlation between attenuation curve slope and dust column density is present, despite of having large scatters, these different methods should give consistent results, which is not seen from those results mentioned above.

In addition, theoretical studies with radiative transfer models often predict a consistent relation between attenuation curve slope and $\av$ that greyer attenuation curves are coupled with higher dust opacities \citep{Witt2000, Chevallard2013, Seon2016, Narayanan2018, Trayford2020, Shen2020, Salim2020}. The origin of this relationship is the increasing contribution of scattering at lower $\av$ \citep{Chevallard2013}. However, these radiative transfer predictions depend on the adopted dust-stars distribution geometry. For instance, \citet{Lin2021} found this relationship exists in a well-mixed geometry but is weak or absent if a two-layer geometry is assumed \citep[see also][]{Witt2000}. Applying the Empirical Dust Attenuation framework to the large-scale cosmological hydrodynamical simulations (i.e. without radiative transfer effects), \citet{Hahn2021} found that the attenuation curve slope (parameterized by $A_{\rm 1500}/\av$) flattens with increasing $\av$. Given that the physical processes related to dust attenuation are complicated, more efforts are demanded to improve the theoretical modelling of the radiative transfer effects (scattering and absorption) in combination with local geometry effects of the interstellar dust in galaxies, in order to reconcile the theoretical predictions with the observational results.

\section{Summary} \label{sec7}

Using the publicly-available FUV to FIR data, we obtained high-quality SEDs for a sample of 2\,291 SFGs selected from the GAMA survey.  We carried out SED fitting for our sample SFGs using CIGALE with model SEDs generated from the well-designed declining starburst SFHs, and derived attenuation parameters and examined the effects of the fitting degeneracies between the attenuation curve slope ($\delta$), dust column density ($\sim\av$) and the intrinsic UV slope ($\beta_0$). Our main findings are summarized as follows:

\begin{enumerate}[1.]

\item The local SFGs exhibit a wide range of $\beta_0$ over [$-$2.4, $-$1.7] with a median of $-$2.0. Modelling of galaxy SEDs with model templates from a constant starburst SFH scan a limited range of $\beta_0<-2.2$, which will significantly bias the determination of attenuation parameters. \\

\item On average, our local SFGs have attenuation curves slightly steeper than the Calzetti curve with $\delta\approx-0.14$. The variations in best-fitting  $\delta$ are dominated by the fitting errors. \\ 

\item There is a strong degeneracy between $\beta_0$ and $\delta$ in SED fitting, i.e. the $\delta$--$\beta_0$ degeneracy. The current SED fitting algorithm is not able to break this degeneracy and constrain the attenuation curve slope well.  \\

\item We find $\av$ is strongly degenerate with $\delta$, which strongly biases the measured $\av$--$\delta$ relation. Instead, $\afuv$ is strictly constrained by the IR luminosity in terms of the energy balance and shows little or no degeneracy with $\delta$. It better measures the dust attenuation than $\av$. \\ 

\item We find the relation between $\delta$ and $\av$ derived from SED fitting is governed by the systematic biases raised by the fitting degeneracies, but does not reflect an intrinsic connection between the two quantities.  \\ 

\item The relatively small degeneracy errors given in \citetalias{Salim2020} are attributed to the use of model SED templates generated with constant starburst SFHs in their SED fitting. The model SEDs appear similarly blue in the UV colour ($\beta_0<-2.2$), and bias the fitting results towards steeper attenuation curves, smaller degeneracy errors, and a stronger $\av$--$\delta$ correlation. \\

\end{enumerate}

While the relation between attenuation curve slope and dust column density (e.g. $\av$) has been widely explored via the SED fitting algorithm, our principal contribution is to demonstrate that this relation likely finds its origin in the systematic biases driven by fitting degeneracies and chosen model SEDs with biased $\beta_0$. We thus argue that the $\av$--$\delta$ correlation derived from SED fitting, i.e. flatter dust attenuation curves being tightly linked with higher dust attenuation in star-forming galaxies, is no longer valid. More efforts, particularly on the accurate determination of dust attenuation curves, are demanded in the future. Our findings are also useful in guiding interpretations of other fitted parameters in SED studies, which are often degenerate to some level. A hierarchical Bayesian approach may help to break these degeneracies in SED fitting if we have rich multi-wavelength datasets, as demonstrated in some previous studies\citep[e.g.,][]{Kelly2012,Juvela2013,Galliano2018a,Lamperti2019}.

\section*{acknowledgments}
We are grateful to the anonymous referee for helpful comments and suggestions that significantly improved the quality of the manuscript. This work is supported by the National Key R\&D Program of China (2017YFA0402703), the National Science Foundation of China (12073078 and 11773076), the Major Science and Technology Project of Qinghai Province (2019-ZJ-A10), the science research grants from the China Manned Space Project with NO. CMS-CSST-2021-A02, CMS-CSST-2021-A04 and CMS-CSST-2021-A07, and the Chinese Academy of Sciences (CAS) through a China-Chile Joint Research Fund (CCJRF \#1809) administered by the CAS South America Centre for Astronomy (CASSACA). SW acknowledges support from the Chinese Academy of Sciences President’s International Fellowship Initiative (grant no. 2022VMB0004). 

GAMA is a joint European-Australasian project based around a spectroscopic campaign using the Anglo-Australian Telescope. The GAMA input catalogue is based on data taken from the Sloan Digital Sky Survey and the UKIRT Infrared Deep Sky Survey. Complementary imaging of the GAMA regions is being obtained by a number of independent survey programmes including GALEX MIS, VST KiDS, VISTA VIKING, WISE, Herschel-ATLAS, GMRT and ASKAP providing UV to radio coverage. GAMA is funded by the STFC (UK), the ARC (Australia), the AAO, and the participating institutions. The GAMA website is \url{http://www.gama-survey.org/}.

\section*{Data Availability}
The data underlying this article will be shared on reasonable request to the corresponding author.

\bibliographystyle{mnras}
\bibliography{references}
\appendix
% \onecolumn

\section{Additional figures}

\begin{figure*}%[htb]
\centering
\includegraphics[width=0.9\textwidth]{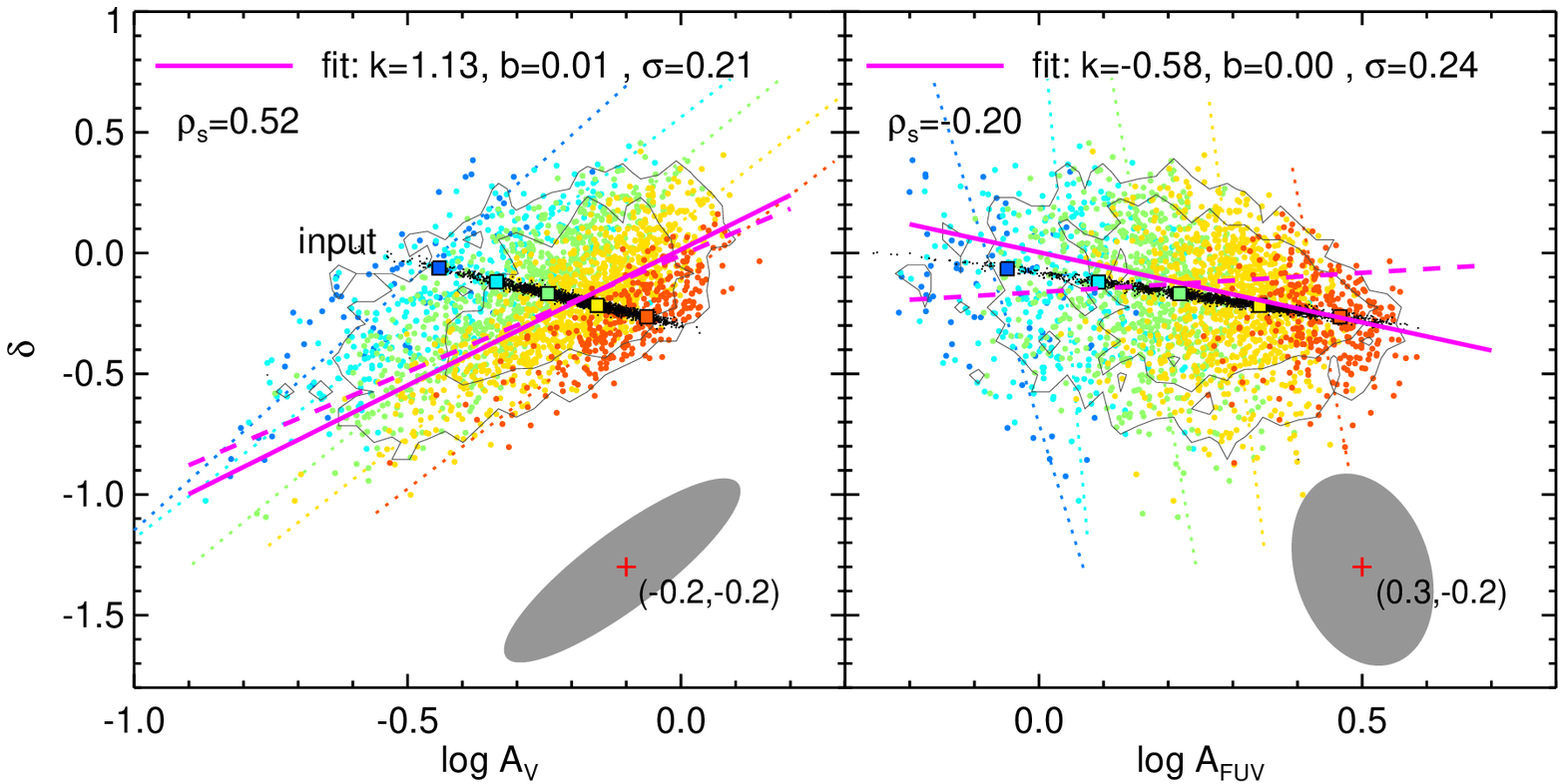}
\includegraphics[width=0.9\textwidth]{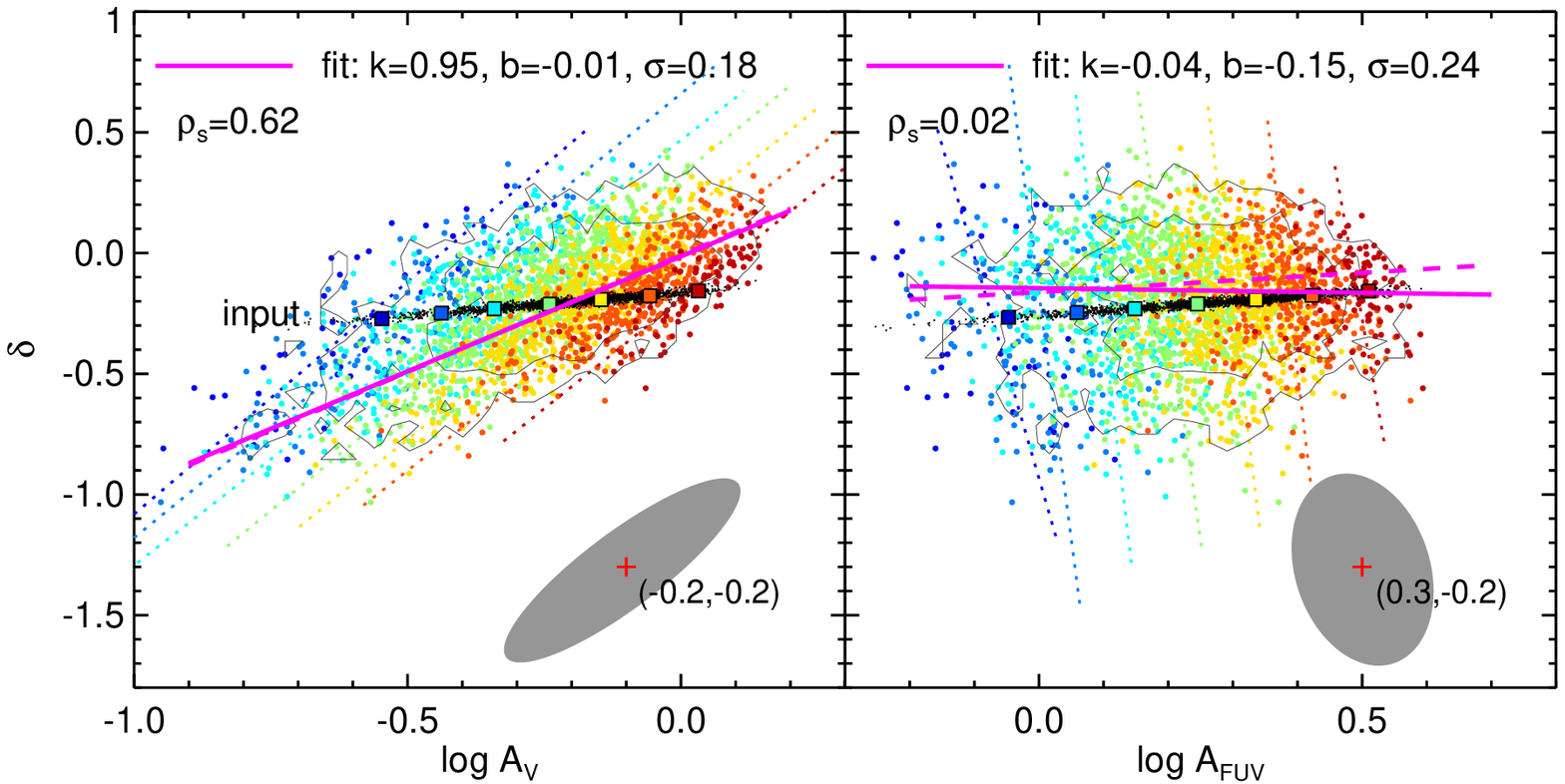}
\caption{Similar to Figure \ref{fig_ax_delta_mock} but showing the results of simulated galaxies having a negative (top, slope $\sim$$-0.5$) and positive (bottom, slope $\sim$0.2) input $\av$--$\delta$ relation, respectively. }
\label{fig_a1}
\end{figure*}

\bsp	% typesetting comment
\label{lastpage}
\end{document}